\documentclass[%
 preprint,
 amsmath,amssymb,
 aps,
]{revtex4-2}

\usepackage{graphicx}
\usepackage{bm}

\usepackage{amsmath}
\newcommand{\norm}[1]{\left\lVert#1\right\rVert}

\newcommand{\bs}{\boldsymbol}

\newcommand{\BE}{\begin{enumerate}}
\newcommand{\EE}{\end{enumerate}}
\newcommand{\BI}{\begin{itemize}}
\newcommand{\EI}{\end{itemize}}

\begin{document}


\title{Using Experimentally Calibrated Regularized Stokeslets to Assess Bacterial Flagellar Motility Near a Surface}
\author{Orrin Shindell}
\affiliation{Trinity University, San Antonio, TX, USA}
 \email{oshindel@trinity.edu}

\author{Hoa Nguyen}
\affiliation{Trinity University, San Antonio, TX, USA}
 \email{hnguyen5@trinity.edu}
 
 \author{Nicholas Coltharp}
\affiliation{Trinity University, San Antonio, TX, USA}
 \email{ncolthar@trinity.edu}
 
  \author{Frank Healy}
\affiliation{Trinity University, San Antonio, TX, USA}
 \email{fhealy@trinity.edu}

   \author{Bruce Rodenborn }
\affiliation{Centre College, Danville, KY, USA}
 \email{bruce.rodenborn@centre.edu}

\date{\today}

\abstract{The presence of a nearby boundary is likely to be important in the life cycle and evolution of motile flagellate bacteria. This has led many authors to employ numerical simulations to model near-surface bacterial motion and compute hydrodynamic boundary effects. A common choice has been the method of images for regularized Stokeslets (MIRS); however, the method requires discretization sizes and regularization parameters that are not specified by any theory. To determine appropriate regularization parameters for given discretization choices in MIRS, we conducted dynamically similar macroscopic experiments and fit the simulations to the data. In the experiments, we measured the torque on cylinders and helices of different wavelengths as they rotated in a viscous fluid at various distances to a boundary. We found that differences between experiments and optimized simulations were less than 5\% when using surface discretizations for cylinders and centerline discretizations for helices. Having determined optimal regularization parameters, we used MIRS to simulate an idealized free-swimming bacterium constructed of a cylindrical cell body and a helical flagellum moving near a boundary. We assessed the swimming performance of many bacterial morphologies by computing swimming speed, motor rotation rate, Purcell's propulsive efficiency, energy cost per distance, and a new metabolic energy cost defined to be the energy cost per body mass per distance. All five measures predicted the same optimal flagellar wavelength independently of body size and surface proximity. Although the measures disagreed on the optimal body size, they all predicted that body size is an important factor in the energy cost of bacterial motility near and far from a surface. 

}

}

\maketitle

\section{Introduction}
\label{sec:intro} 

Living organisms emerge, evolve, and reside within habitats, and the physical interactions among organisms and their environments impose selective forces on their evolution. In their low Reynolds number surroundings, bacteria such as \textit{Escherichia coli} and \textit{Pseudomonas aeruginosa} have evolved a mechanical motility system to propel themselves through fluids. This system consists of one or more helical flagella, and these flagellar organelles are attached to the cell body by rotary nanomotors.
Flagellar motor rotation is driven by an ion flow through the motor, causing the flagellum and the bacterial cell body to rotate in opposite directions \cite{sowa2008bacterial}. 
A bacterium swimming through a fluid can be described as a non-inertial system in which the mechanical power output by the motor is instantaneously dissipated by fluid drag on the body and flagellar filaments. The interaction between the bacterium and the fluid generates a flow that results in the net motion of the bacterium. 
Different flows can be more or less favorable to the survival of an organism \cite{lauga2016bacterial}; and the presence of a surface introduces boundary effects that modify how a swimming cell interacts with the fluid. We consider here the example of a unicellular motile flagellate bacterium swimming through a fluid near to a surface and how the conformation of the bacterial cell body and the flagellar organelle may be optimized for such an environment.

The efficiency of the bacterial motility system has been the focus of numerous theoretical \cite{purcell1977life, higdon1979hydrodynamics, shapere1989efficiencies}, computational \cite{shum2010modelling, spagnolie2011comparative, acemoglu2014effects, he2014propulsive, bet2017efficient,schuech2019motile}, and experimental works \cite{chattopadhyay2006swimming, li2006low, jeon2012flow}. 
In an early paper on swimming efficiency, E. Purcell discussed two measures: the propulsive efficiency (Purcell efficiency) and the energy consumed during bacterial motion per body mass \cite{purcell1977life}. 
The Purcell efficiency--a specialized form of the Lighthill efficiency \cite{lighthill1952squirming} for rotary motor-driven bacterial propulsion--is defined as the ratio of the least power needed to translate a bacterial body against fluid drag to the total power output by the motor during motion of the bacterium. 
Most work has focused on the Purcell efficiency because it is a scale-independent function of the geometries of the cell body and flagellum. 
One shortcoming of this measure, however, is that it is independent of the motor's response to an external load imposed by the environment and therefore cannot assess the biological fitness of the bacterial motor.
Another measure of bacterial performance used by a few authors is the distance traveled by a bacterium per energy input by the motor \cite{li2006low, li2017flagellar}, which provides a different means of evaluating fitness, as explained below.

In this work, we investigate and compare predictions of the optimal bacterial motility system made by five measures. The first two measures are related directly to the motion of a bacterium: the swimming speed and the motor rotation frequency. Bacteria live in an environment where nutrients diffuse on time and length scales comparable to bacterial motion. To effectively achieve chemotaxis, bacteria must move quickly enough to sample their chemical environment before it is randomized by diffusion \cite{purcell1977life,schuech2019motile}. The bacterial motor has a characteristic frequency response that depends on the external torque load \cite{chen2000torque,sowa2003torque,xing2006torque,Darnton2007flagparameters}. At low frequencies, small changes in applied load correspond to large changes in operating frequency, whereas at high frequencies, small changes in load correspond smaller changes in frequency. In the low speed regime, the motion may be unreliable because small changes in applied load that occur, for example, by approaching a boundary could lead to the motor stalling. However, the low speed regime is more thermodynamically efficient than the high speed regime. These two competing effects must be balanced to achieve a strong swimming performance.

The other three performance measures we studied are based on the mechanical energy cost to achieve motility: the Purcell inefficiency (or the inverse of the Purcell efficiency), the inverse of distance traveled per energy input, and the metabolic energy cost, which we define to be the energy output by the motor per body mass per distance traveled.
Each of these measures compares the ratio of the power output of the bacterial motor to the performance of a particular task. 
The rationale for introducing the metabolic cost function is that it measures the actual energetic cost to the organism to perform a specific biologically relevant task, i.e., translation through the fluid. Moreover, the metabolic energy cost depends upon the rotation speed of the motor and, because the bacterial motor has a different responses to different external conditions, predicts different optimal morphologies based on the environment than the other measures. 

To determine the values of performance measures attained by different bacterial geometries, we employed the method of regularized Stokeslets \cite{cortez2001method} and the method of images for regularized Stokeslets (MIRS) \cite{ainley2008method}, which includes the effect of a solid boundary. Employing MRS and MIRS requires determining values for two kinds of free parameters: those associated with computation and those associated with the biological system. As with any computational method, the bacterial structure in the simulation is represented as a set of discrete points. The body forces acting at those points are expressed as a vector force multiplied by a regularized distribution function, whose width is specified by a regularization, or ``blob'' parameter. 
Though other simulations have produced numerical values for dynamical quantities like torque \cite{das2018computing} that are within a reasonable range for bacteria, precise numbers are not possible without an accurately calibrated method.

There is no known theory that predicts the relationship between the discretization and regularization parameters, though one benchmarking study showed that MRS simulations could be made to match the results of other numerical methods \cite{martindale2016blobsize}. To determine the optimal regularization parameter for chosen discretization sizes,
we performed  dynamically similar macroscopic experiments using the two objects from our model bacterium: a cylinder and a helix, see Fig. \ref{fig:schematic}.
Such an approach was previously used to evaluate the accuracy of various computational and theoretical methods for a helix \cite{rodenborn2013propulsion}. 
By measuring values of the fluid torque acting on rotating cylinders near a boundary, we verified the theory of Jeffery and Onishi \cite{jeffrey1981slow}, which in turn we used to calibrate the ratio of discretization to regularization size in MRS and MIRS simulations of rotating cylindrical cell bodies.  
For helices there are no exact analytical results. To determine regularization parameters for helices we descretized them along their centerlines and fit simulation results directly to experimental measurements. Calibrating our simulations of rotating cylinders and helices with the experiments allowed us to build a bacterial model with a cylindrical cell body and a helical flagellum whose discretization and regularization parameter are optimized for each part.

To impose motion on the bacterial model, we needed only to specify the motor rotation -- a consequence of there being no body forces acting on the bacterium \cite{das2018computing}.  
The motor rotation rate, however, depends upon the external load \cite{chen2000torque, sowa2003torque, xing2006torque, li2006low}.  In our simulations, we ensured the motor rotation rate and the total torque acting on the motor match a point on the experimentally determined torque-speed response curve reported in the literature \cite{chen2000torque,Darnton2007flagparameters}. 
The dynamical quantities output from the simulations were then used to compute performance measures for different bacterial geometries at various distances from the boundary.

Our paper is organized as follows: Sec. \ref{sec:methods} discusses our implementation of the MRS
 and the MIRS, our use of dynamically similar experiments to calibrate the simulations, and our determination of the torque-speed response curve for the motor; Sec. \ref{sec:results} compares our five fitness measures: free swimming speed, motor frequency, inverse Purcell efficiency, energy per distance and metabolic cost per distance; and Sec. \ref{sec:discussion} discusses the predictions made by each fitness measure and comments on future directions of our work. 

\section{Materials and Methods}\label{sec:methods}

\subsection{Numerical Methods}
\begin{figure}[ht] 
\centering
\includegraphics[width=\textwidth]{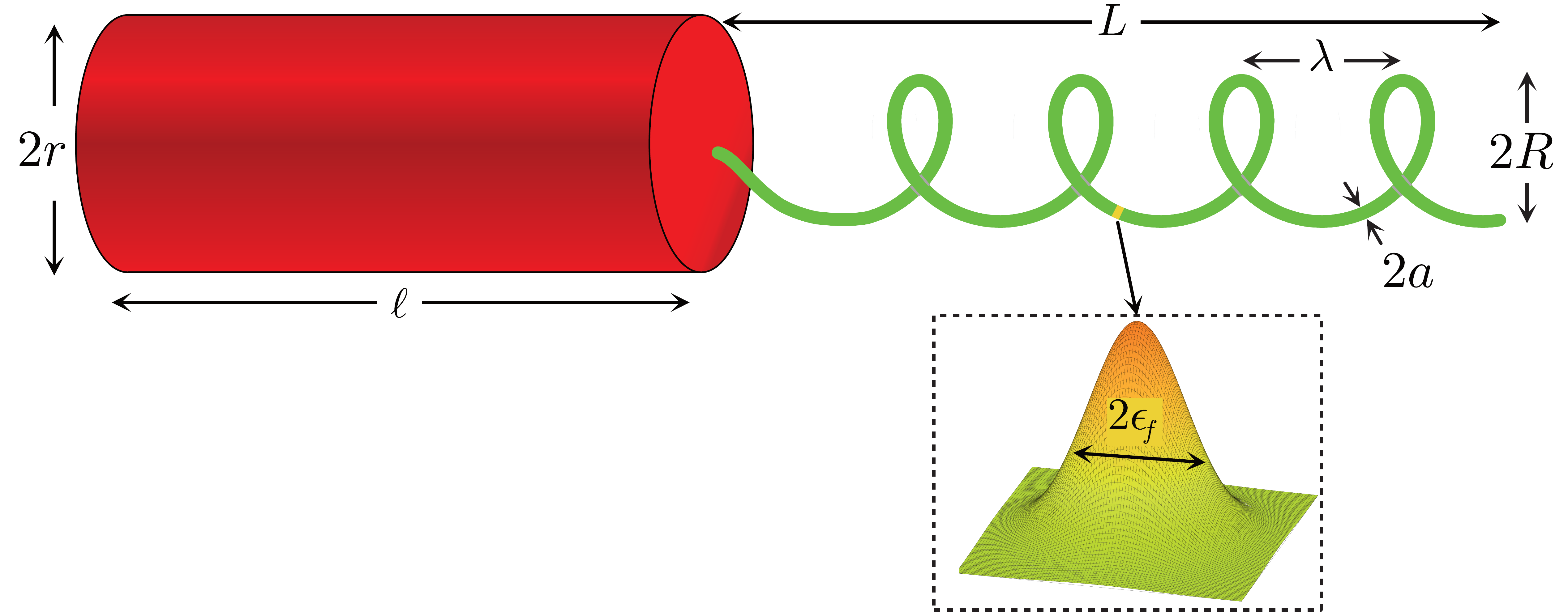}
\caption{Schematic of our model bacterium with flagellar radius $R$, wavelength $\lambda$, axial length $L$, and filament radius $a$.  
The body of the bacterium was modeled as a cylinder with radius $r$ and length $\ell$. Each flagellum was modeled as a regular helix that tapers to zero radius at the point it attaches to the body. Our simulations used a surface discretization of regularized Stokeslets to represent the cylinder
and a string of regularized Stokeslets along the centerline of the flagellum.
The inset represents a radially symmetric blob function described in Sec. \ref{sec:MRS} that is used to spread the force at a given point on the flagellar centerline. For the purpose of illustration, we show the blob function of two variables whose width is controlled by the regularization parameter $\epsilon_f$.
\label{fig:schematic}}
\end{figure}

Bacterial motility using a helical flagellum often involves multiple flagella, and bodies may be spherical, cylindrical, or helical \cite{young2006selective}. 
We reduced the complexity by considering a simpler biomechanical system of a regular cylindrical body to which a single, uniform flagellum is attached, as shown in Fig. \ref{fig:schematic}. 
This simple system, however, contains the same essential geometric factors as some real bacteria such as \textit{E. coli}, which have a long rod-shaped body and helical flagella that bundle together, forming a single helix.
Our goal was to assess how the performance of our  model organism changes when its geometrical parameters and distance to an infinite plane wall are varied in numerical simulations. 
We quantified the performance of different models by computing speed, motor rotation rate, and the 
three energy cost measures.

\begin{figure}[ht] 
\centering
\includegraphics[width=\textwidth]{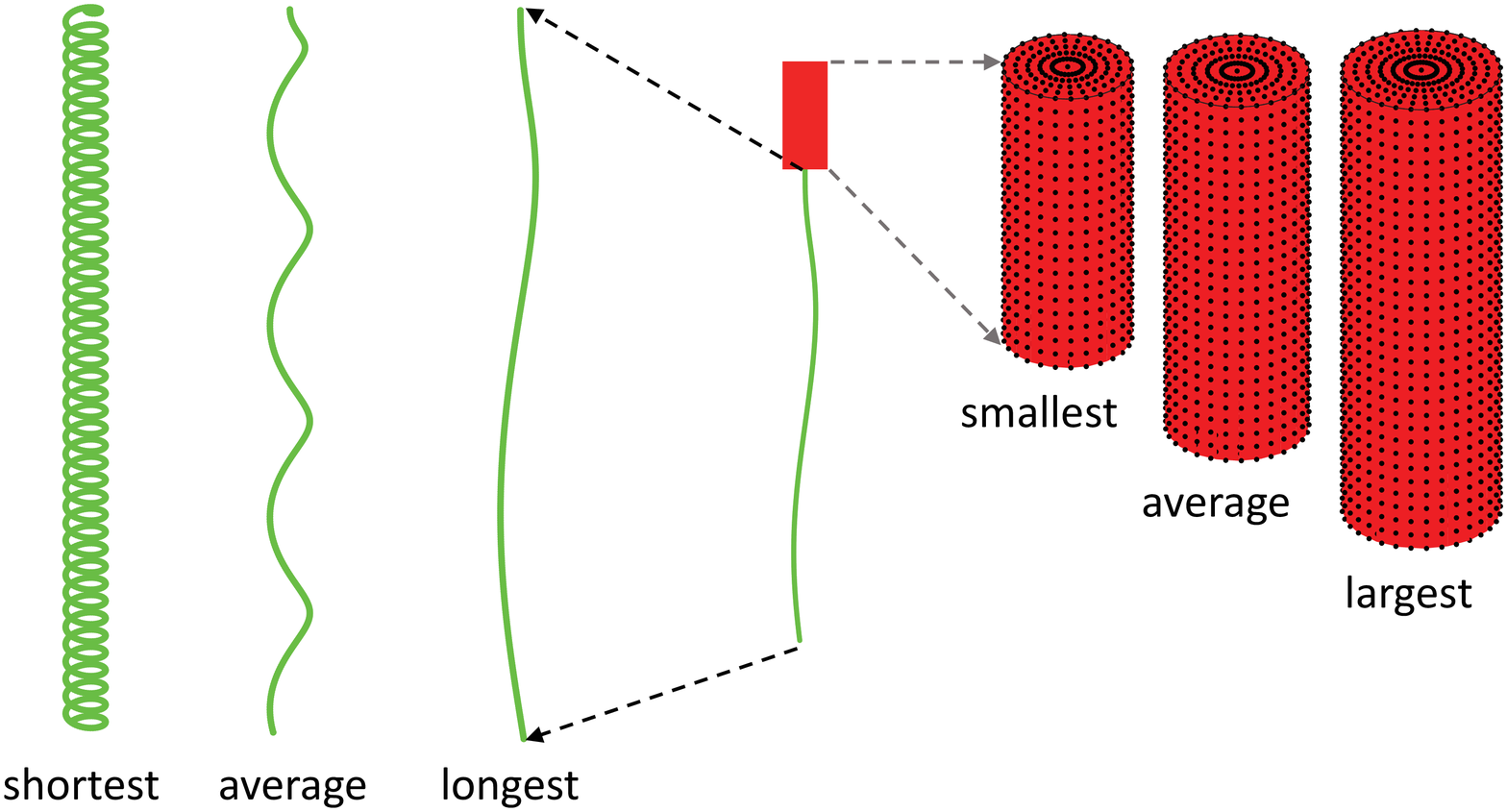}
\caption{Our model bacterium had a cylindrical cell body and a helical flagellum, and 25 different cell body sizes and eighteen different flagellar wavelengths were used, as described in Table \ref{Table_bac_model}. 
Three cell bodies with the smallest, average, and largest volumes, respectively are shown on the right whereas the three flagella with the shortest, average, and longest wavelengths are presented on the left. The middle shows an example of one such model, which has the smallest body and the longest wavelength flagellum.}
\label{fig:bacterium_models}
\end{figure}

We composed our model of a bacterium with a cylindrical cell body and a tapered left-handed helical flagellum as shown in Fig. \ref{fig:schematic} and Fig. \ref{fig:bacterium_models}. 
The flagellar centerline is described by
\begin{equation}\label{Eq_flag}
\left\{
\begin{aligned}
 x(s) &= (1 - e^{-k^2 s^2})R\sin(ks + \theta)\\
 y(s) &= (1 - e^{-k^2 s^2})R\cos(ks + \theta)\\
 z(s) &= s
\end{aligned} 
\right.
\end{equation}
where $0 \le s \le L$ and $L$ is the axial length in the $z$-direction. 
$k$ is the wavenumber $2\pi/\lambda$ where $\lambda$ is the wavelength. 
$\theta$ is the phase angle of the helical flagellum at 16 evenly spaced phases. 

The parameter values used for the bacterium models shown in Fig. \ref{fig:bacterium_models} are given in Table \ref{Table_bac_model}.

\begin{table}[ht] 
\caption{Parameters used in numerical simulations.} 
\label{Table_bac_model}
\begin{tabular}{lcccc}
\toprule
\textbf{Parameter}	& \textbf{Symbol}	& \textbf{Value} & \textbf{Unit} &  \textbf{Reference}\\
\hline
Dynamic viscosity of the fluid  &$\mu$  &$0.93$ &$cP$ & \\
\hline
\textbf{Cell body (Cylinder)} & & & & \\
length & $\ell$ &(a) &$\mu m$ &\cite{Darnton2007flagparameters}\\
radius &$r$ &(b) &$\mu m$ &\cite{Darnton2007flagparameters} \\
optimal discretization factor &$\gamma_c$  &$6.4$  &   &  \\
discretization size &$ds_c$  &$0.096$   &$\mu m$  &  \\
regularization parameter &$\epsilon_c  = ds_c / \gamma_c$  &$0.015$      &$\mu m$    &   \\
\hline
\textbf{Flagellum (Helix)} & & & & \\
axial length &$L$ &$8.3$ &$\mu m$ &\cite{Darnton2007flagparameters} \\
wavelength &$\lambda$ &(c) &$\mu m$ &\\
helix radius &$R$ &$0.2$ &$\mu m$ &\cite{Darnton2007flagparameters} \\
filament radius &$a$  &$0.012$ &$\mu m$ &\cite{Darnton2007flagparameters} \\
initial motor frequency &$\Omega_m/(2\pi)$ &$154$ &$Hz$ &\cite{Darnton2007flagparameters} \\
optimal filament factor &$\gamma_f$  &$2.139$  &   &  \\
regularization parameter &$\epsilon_f = \gamma_f a$  &$0.026$      &$\mu m$    &   \\
discretization size &$ds_f = \epsilon_f$  &$0.026$   &$\mu m$  &  \\
\hline
\shortstack{distance from flagellar axis to wall} &$d$ &(d) &$\mu m$ & \\
\hline

\end{tabular}
\footnotesize{\newline 
(a) $\ell \in$ \{1.9, 2.2, 2.5, 2.8, 3.1\} $(\mu m)$\\
(b) $r \in$ \{0.395, 0.4175, 0.44, 0.4625, 0.485\} $(\mu m)$\\
(c) $\lambda \in$ \{0.2, 0.5, 0.8, 1.1, 1.4, 1.7, 2.02, 2.22, 2.3, 2.42, 2.6, 2.9, 3.2, 3.6, 4.0, 5.0, 7.0, 9.0 \} $(\mu m)$\\
(d) $d \in$ \{0.55, 0.62, 0.71, 0.82, 0.96, 1.12, 1.32, 1.56, 1.85, 2.20, 2.26, 2.52, 2.81, 3.14, 3.5, 3.93, 4.4, 4.93, 5.53, 6.2, 8.2, 10.2\} $(\mu m)$
}
\end{table}

\subsubsection{Method of regularized Stokeslets} \label{sec:MRS}
The microscopic length and velocity scales of bacteria ensure that fluid motion at that scale can be described using the incompressible Stokes equations. 
We used the MRS in three dimensions \cite{cortez2001method} to compute the fluid-bacterium interactions due to the rotating flagellum in free space at steady state:
\begin{equation} \label{Eq_MRS}
\begin{split}
\mu  \triangle \mathbf{u(\mathbf{x})} - \nabla p(\mathbf{x}) & = -\mathbf{F(\mathbf{x})} \\
\nabla \cdot \mathbf{u(\mathbf{x})} & = 0
\end{split}
\end{equation}
$\mathbf{u}$ is the fluid velocity, $p$ is the fluid pressure, and $\mu$ is the dynamic viscosity.
$\mathbf{F}$ is the body force represented as $\mathbf{f}_k\phi_\epsilon (\mathbf{x} - \mathbf{x}_k)$ where $\mathbf{f}_k$ is a point force at a discretized point $\mathbf{x}_k$ of the bacterium model. 
In our simulations, we used the blob function $\phi_\epsilon (\mathbf{x} - \mathbf{x}_k)= \frac{15 \epsilon^4}{8\pi(r_k^2 + \epsilon^2)^\frac{7}{2}}$ where $r_k = \norm{\mathbf{x} - \mathbf{x}_k}$.
This radially symmetric smooth function depends on a regularization parameter $\epsilon$ which controls the spread of the point force $\mathbf{f}_k$.
Given $N$ such forces, the resulting velocity at any point $\mathbf{x}$ in the fluid can be computed as
\begin{equation} \label{Eq_MRS_velocity}
\mathbf{u}(\mathbf{x}) = \frac{1}{8\pi\mu}\sum_{k = 1}^N \frac{\mathbf{f}_k(r_k^2 + 2\epsilon^2)}{(r_k^2 + \epsilon^2)^\frac{3}{2}} + \frac{(\mathbf{f}_k \cdot (\mathbf{x} - \mathbf{x}_k))(\mathbf{x} - \mathbf{x}_k)}{(r_k^2 + \epsilon^2)^\frac{3}{2}} = \frac{1}{8\pi\mu}\sum_{k = 1}^N S_\epsilon(\mathbf{x},\mathbf{x}_k)\mathbf{f}_k 
\end{equation}
Evaluating Eq. \ref{Eq_MRS_velocity} $N$ times, once for each $\mathbf{x}_k$, yields a $3N\times3N$ linear system of equations for the velocities of the model points. 
In the limit as $\epsilon$ approaches 0, the resulting velocity $\mathbf{u}$ approaches the classical singular Stokeslet solution. 
In practice, the specific choice of $\epsilon$ may depend on the discretization or the physical thickness of the structure.

In our bacterium model, we discretized the cell body as $N_c$ points on the surface of a cylinder, and we modeled the flagellum as $N_f$ points distributed uniformly along the arclength of the centerline.
In Sec. \ref{sec:results}, we present the optimal regularization parameter for the cylindrical cell we obtained by calibrating the simulations based on the experiments and theory. 
The regularization parameter for the helical flagellum was found by calibrating simulations with experiments, since there is no exact theory for rotating helices, as presented in Sec. \ref{sec:expts}.

\subsubsection{Method of images for regularized Stokeslets}\label{sec:MIRS}
We used the method of images for regularized Stokeslets (MIRS)\cite{ainley2008method} to solve the incompressible Stokes equations (Eq. \ref{Eq_MRS}) and simulate bacterial motility near a surface. 
In the method, the no-slip boundary condition on an infinite plane wall is satisfied by imposing a combination of a Stokeslet, a Stokeslet doublet, a potential dipole, and rotlets at the image point $\mathbf{x}^*_k$ of each discretized point $\mathbf{x}_k$.
The image point $\mathbf{x}^*_k$ is the point obtained by reflecting $\mathbf{x}_k$ across the planar surface. 
The resulting velocity at any point $\mathbf{x}$ in the fluid bounded by a plane can be found in Ref.\cite{ainley2008method} and written in the compact form similar to Eq. \ref{Eq_MRS_velocity}:

\begin{equation} \label{Eq_MIRS_velocity}
\mathbf{u}(\mathbf{x}) = \frac{1}{8\pi\mu}\sum_{k = 1}^N S^*_\epsilon(\mathbf{x},\mathbf{x}_k)\mathbf{f}_k 
\end{equation}

\subsubsection{Force-free and torque-free models}
On a free-swimming bacterium, the only external forces acting are due to the fluid-structure interaction. A bacterium is a non-inertial system so the net external force and net external torque acting on it must vanish.
This means that $\mathbf{F}_c+ \mathbf{F}_f = \mathbf{0}$ and $\bs{\tau}_c + \bs{\tau}_f = \mathbf{0}$, where $\mathbf{F}_c$ / $\bs{\tau}_c$ and $\mathbf{F}_f$ / $\bs{\tau}_f$ represent, respectively, the net fluid forces and torques acting on the cell body and flagellum. These force-free and torque-free constraints 
require the cell body and flagellum to counterrotate relative to each other.  In our simulations, the point connecting the cell body and the flagellum $\mathbf{x}_r$ represented the motor location, and was used as the reference point for computing torque and angular velocity.

Given an angular velocity $\mathbf{\Omega}_m$ of the motor, the relationship between the lab frame angular velocities of the flagellum and the cell body is $\mathbf{\Omega}_f = \mathbf{\Omega}_c + \mathbf{\Omega}_m$ \cite{das2018computing}. 
Since $\mathbf{\Omega}_m$ is the relative rotational velocity of the flagellum with respect to the cell body, the resulting velocity $\mathbf{\Tilde{u}}(\mathbf{x}_k)$ at a discretized point $\mathbf{x}_k$ on the flagellum ($k = 1,...,N_f$) can be computed as $\mathbf{\Omega}_m \times \mathbf{x}_k$ (this velocity is set to zero at a discretized point on the cell body). 
Using the MRS (or MIRS) and the six added constraints from the force-free and torque-free conditions, we formed a $(3N + 6) \times (3N + 6)$ linear system of equations to solve for the translational velocity $\mathbf{U}$ and angular velocity $\mathbf{\Omega}_c$ of the cell body and the 
internal
force $\mathbf{f}_k$ 
acting
at 
the
discretized point $\mathbf{x}_k$ of the model:

\begin{equation} \label{Eq_MRS_force_torque_free}
\begin{split}
\mathbf{\Tilde{u}}(\mathbf{x}_j) &= \frac{1}{8\pi\mu}\sum_{k = 1}^N G_\epsilon(\mathbf{x}_j,\mathbf{x}_k)\mathbf{f}_k - \mathbf{U} - \mathbf{\Omega}_c \times (\mathbf{x}_j - \mathbf{x}_r), \quad j = 1,...,N\\ 
\sum_{k = 1}^N \mathbf{f}_k &= \mathbf{0}, \quad\quad
\sum_{k = 1}^N (\mathbf{x}_k - \mathbf{x}_r) \times \mathbf{f}_k = \mathbf{0}
\end{split}
\end{equation}
where $G_\epsilon$ is $S_\epsilon$ from Eq. \ref{Eq_MRS_velocity} for swimming in a free space or $S^*_\epsilon$ from Eq. \ref{Eq_MIRS_velocity} for swimming near a plane wall.
Each $\mathbf{f}_k$ represents a point force acting at point $\mathbf{x}_k$, which is in principle an internal contact force due to interactions with the points on the bacterium that neighbor $\mathbf{x}_k$.
Each $\mathbf{f}_k$ is balanced by the hydrodynamic drag that arises from a combination of viscous forces and pressure forces exerted on the point $\mathbf{x}_k$ by the fluid (Eq. \ref{Eq_MRS}).
 
By
computing each $\mathbf{f}_k$, we were able to deduce the fluid interaction with each point of the bacterial model.
Eq. \ref{Eq_MRS_force_torque_free} shows that the calculated quantities $\mathbf{U}$, $\mathbf{\Omega}_c$, $\mathbf{F}_c$, and $\bs{\tau}_c$ depend linearly on the angular velocity $\mathbf{\Omega}_m$ since $\mathbf{\Tilde{u}}(\mathbf{x}_j) = \mathbf{\Omega}_m \times \mathbf{x}_j$. 
\subsection{Torque-speed motor response curve}\label{sec.torque-speed}
The singly-flagellated bacteria we simulated move through their environment by rotating their motor, which causes their body and flagellum to counter-rotate accordingly. 
Drag force from the fluid exerts equal magnitude torques on the body and the flagellum, and the value of the torque equals the torque load applied to the motor.
The relationship between the motor rotation rate and the torque load is characterized by a torque-speed curve, which has been measured experimentally in several organisms \cite{chen2000torque, sowa2003torque, xing2006torque, li2006low, Darnton2007flagparameters}. 
In the context of motor response characteristics, speed refers to frequency of rotation. 
We estimated the torque-speed curve for \textit{E. coli} with typical values taken from the literature  \cite{chen2000torque, Darnton2007flagparameters} to match the body and flagellum parameters also taken from measurements on \textit{E. coli} \cite{Darnton2007flagparameters}.

\begin{figure}[ht] 
\centering
\includegraphics[width=\textwidth]{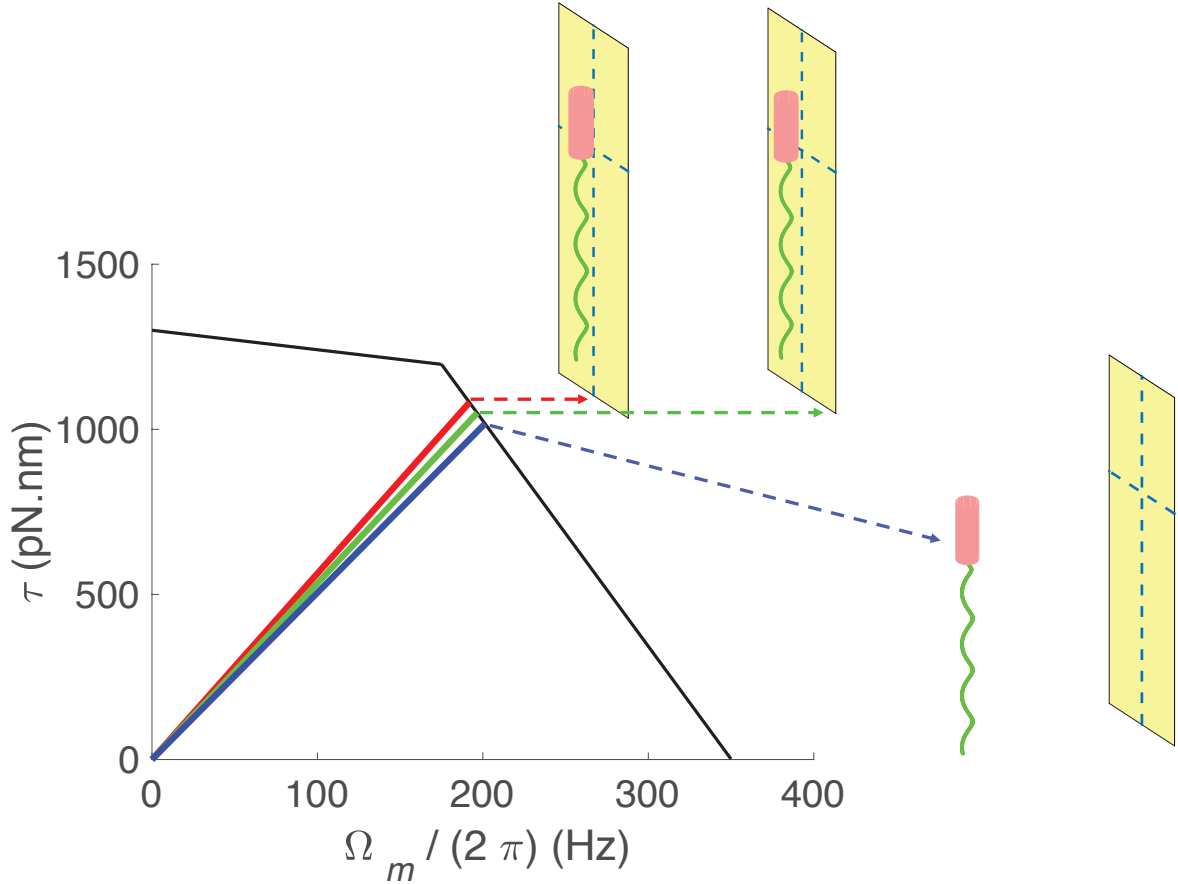}
\caption{Illustration of the estimated torque-speed curve for \textit{E. coli} \cite{chen2000torque,Darnton2007flagparameters}. 
There are two operating regimes: a relatively flat low speed regime $0\leq \Omega_m/2\pi\le175$ Hz where the torque drops from its maximum value of 1300 pN$\cdot$nm at 0 Hz to 1196 pN$\cdot$nm at 175 Hz and a relatively steep high speed regime $175 \leq \Omega_m/2\pi\leq 350$ Hz where the torque drops from 1196 pN$\cdot$nm at 175 Hz to 0 pN$\cdot$nm at 350 Hz.
The insets depict a bacterium model with the average body length $\ell= 2.5$ $\mu$m, the smallest body radius $r = 0.395$ $\mu$m, and the average flagellar wavelength $\lambda = 2.22$ $\mu$m at different distances from the boundary: $d = 8.2$ $\mu$m (blue), $d = 0.71$ $\mu$m (green), $d = 0.54$ $\mu$m (red).  
At closer distances the torque versus rotation rate load lines are steeper so that they intersect the torque-speed curve at a slower rotation speed.}
\label{fig:torque_speed_curve}
\end{figure}

The fluid torque exerted on a rotating object is proportional to its rotation rate under constant environmental conditions in Stokes flow, and thus
plotting the fluid torque versus rotation rate in fixed conditions yields a straight line.
Fig. \ref{fig:torque_speed_curve} shows examples of these `load lines' computed
for our bacterial model at different distances from the boundary; the shallower blue line is calculated for a bacterium far from the boundary, and the steeper red line is calculated near the boundary. 
The load lines shown in Fig. \ref{fig:torque_speed_curve} were computed with typical body and flagellum parameters for \textit{E. coli} \cite{Darnton2007flagparameters}.

The torque-speed curve of the \textit{E. coli} motor has been determined experimentally by measuring the rotation rate of a bead attached to a flagellar stub and then computing the torque on the bead due to fluid drag.
By performing the measurement in fluids of different viscosities, many points on the torque-speed curve were assembled. 
It was found that the torque-speed curve of the \textit{E. coli} bacterial motor decreases monotonically from a maximum stall torque (i.e. the zero-speed torque) of about 1300 pN$\cdot$nm to zero torque, which occurs at a maximum speed of 350 Hz \cite{chen2000torque, xing2006torque,Darnton2007flagparameters}.
There are two linear operating regimes, a low speed regime from 0-175 Hz and a high speed regime 175-350 Hz. 
In the low speed regime below 175 Hz, the torque is a relatively flat function of the motor rotation rate, falling to 0.92 of the stall torque at 175 Hz. 
In the high speed regime above 175 Hz, the torque falls steeply to zero at 350 Hz. 
The torque-speed curve is thus expressed as a piecewise linear function of the motor rotation rate, $\Omega_m$:

\begin{equation}
    \tau = 
    \begin{cases}
    \text{$\displaystyle\left(-0.59\left(\frac{\Omega_m}{2\pi}\right) + 1300\right)$ pN$\cdot$nm} &  \text{for $\displaystyle0\le\frac{\Omega_m}{2\pi}\le175$ Hz}\\
    \text{$\displaystyle\left(-6.83\left(\frac{\Omega_m}{2\pi}\right) + 2392\right)$ pN$\cdot$nm} &  \text{for $\displaystyle175\le\frac{\Omega_m}{2\pi}\le300$ Hz}
    \end{cases}
\end{equation}

Fig. \ref{fig:torque_speed_curve} shows the torque-speed curve as a solid black line. 
In each of our simulations, we ensured that the prescribed motor speed and the computed torque load formed a pair that corresponded to a point on that line.

\subsection{Dynamically similar experiments}\label{sec:expts}
Experiments were performed in an 45-liter tank  (300 mm $\times$ 500 mm $\times$ 500 mm high) filled with
incompressible silicone oil (Clearco\textsuperscript{\textregistered}) with density 970 kg/m$^3$ and dynamic viscosity $\mu$ = $1.13\times 10^{2}$ kg/(m$\cdot$s) at $22^\circ$C, about $10^5$ times that of water. 
The length and speed scales in the experiment ensured that the incompressible Stokes equations Eqs. \ref{Eq_MRS} were valid.
The viscosity of the oil drifted from the manufacturer's stated value ($\mu$ = $1.00\times 10^{2}$ kg/(m$\cdot$s)) very slowly over a two-year period, so we determined the modified viscosity by measuring the torque on rotating cylinders at the center of the tank and recorded data within two months of that measurement.  

The theoretical value for torque per unit length on an infinite rotating cylinder in Stokes flow is $\sigma =4\pi \mu \Omega r^2$, where $\mu$ is the dynamic viscosity of the fluid, $\Omega$ is the angular rotation rate, and $r$ is the cylindrical radius.
We measured the torque $\tau$ on a rotating cylinder with radius $r=6.35\pm 0.2$ mm and length $\ell=149\pm 1$ mm and, by assuming $\tau = \ell\sigma$, used the data to solve for the viscosity of the fluid. 
We also assumed that the finite size of the tank did not affect the torque value in the middle, which was more than $20r$ from the nearest boundary. 
Before each data collection run, we measured the temperature of the oil with a NIST-traceable calibrated thermistor (Cole-Parmer Digi-Sense-AO-37804-04 Calibrated Digital Thermometer) and adjusted the previously determined viscosity using the manufacturer's temperature coefficient of viscosity $1.00\times 10^{-6}$ kg/(m$\cdot$s)/$^\circ$C.
See Sec. \ref{sec:cyl_torque} for a detailed description of the torque measurements.


\subsubsection{Fabricating helices}
     
\begin{figure}[ht] 
\centering
\includegraphics[width=\textwidth]{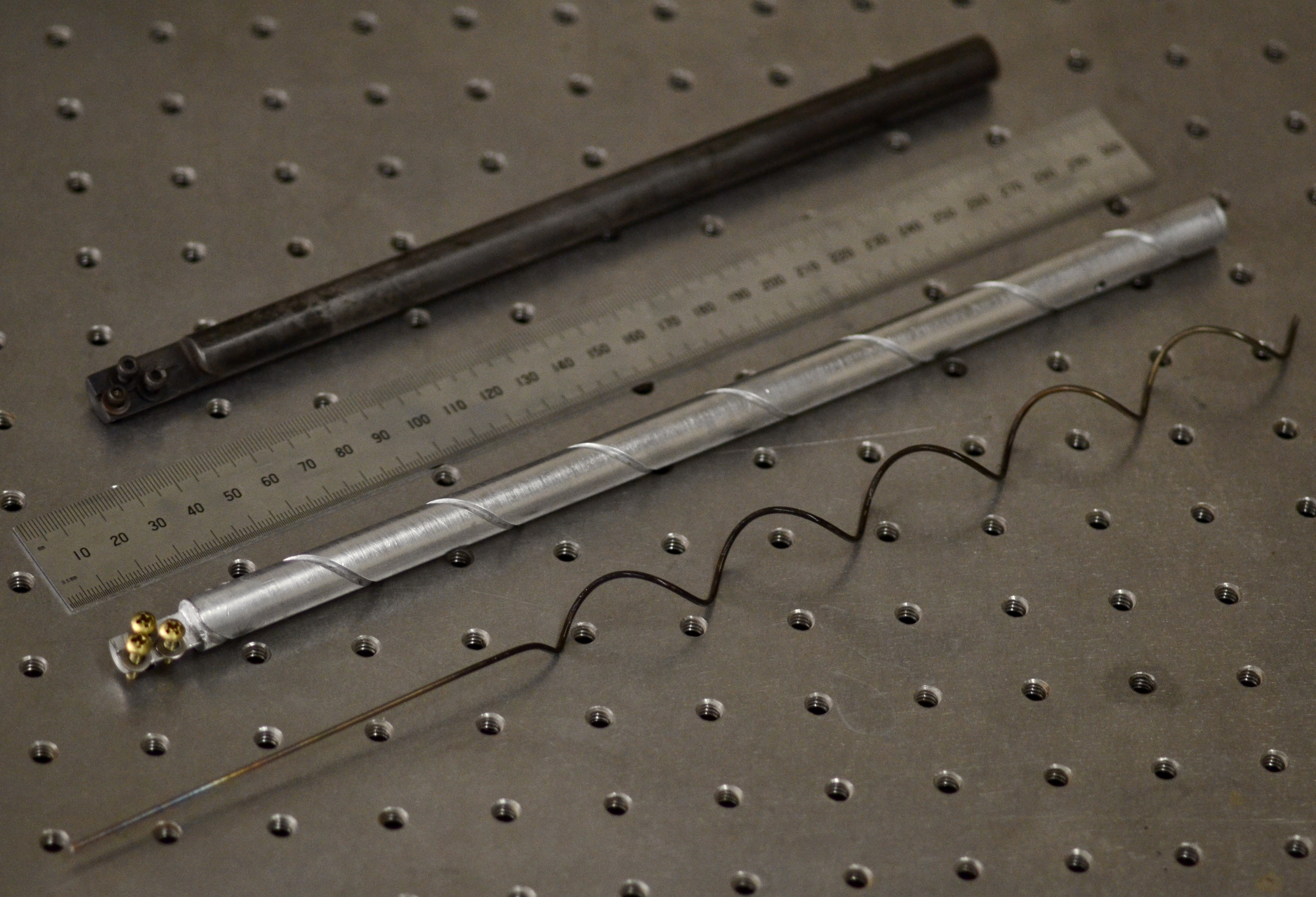}
\caption{Aluminum rods with helical V-shaped grooves were used to form model flagella with different wavelengths.
After forming, the flagella were annealed on a precision rod to increase uniformity in the radius ($\Delta R <0.1mm$). 
The helical parameters are listed in Table \ref{tab:flagella}.}
\label{fig:mandrels}
\end{figure}

We fabricated helices of varying wavelengths ($2.26<\lambda/R<11.88$) by wrapping straight stainless steel welding wire around cylindrical aluminum mandrels with different helical V-grooves precisely machined using a CNC lathe. The V-grooves transition to a flat face with a straight groove,  to which the remaining straight section can be clamped; see Fig. \ref{fig:mandrels}.

Mandrels were held on a lathe, and the wire was hand-spun into the V-groove. The straight sections were secured to the flat faces, which left straight stems aligned with the axes of the helices to be attached to the motor via a rigid shaft adapter. 
Residual tension in the wires caused the wavelengths and radii to vary after they were removed from the mandrels. The helices were forced onto a precision stainless steel rod with radius  $R=6.350\pm0.013$ mm for annealing. The helices on the rod were then placed into a tube furnace (MTI GSL-1500X) and annealed at 900 degrees Celsius for two hours, which removed most of the variation in the radii of the helices and fixed the helical wavelength. 
  
The helix parameters used in the experiments are listed Table \ref{tab:flagella}:

\begin{table}[ht] 
	          
\caption{Wavelengths and lengths of helices.
\label{tab:flagella}}
\begin{tabular}{rr}
\toprule
 \textbf{$\lambda/R$} & \textbf{$L/R$}\\
\hline
	2.26 $\pm$ 0.13		& 22.3 $\pm$ 0.5\\
        3.88 $\pm$ 0.01		& 24.3 $\pm$ 0.5\\
        5.86 $\pm$ 0.08		& 30.0 $\pm$ 0.5\\
        8.65 $\pm$ 0.01		& 23.3 $\pm$ 0.5\\
        10.91 $\pm$ 0.01		& 24.2 $\pm$ 0.5\\
         11.88 $\pm$ 0.01		& 23.1 $\pm$ 0.5\\
\hline
\end{tabular}
\footnotesize{\newline 
The 
helical
wavelength $\lambda$ and axial length $L$ are expressed in terms of the helical radius $R$; $R = 6.35 \pm 0.10$ mm in all cases. 
The filament radius was $a/R= 0.111$ for all helices.
}
\end{table}

\subsubsection{ Axial torque measurements}\label{sec:cyl_torque}
\begin{figure}
    \centering
    \includegraphics[width=\textwidth]{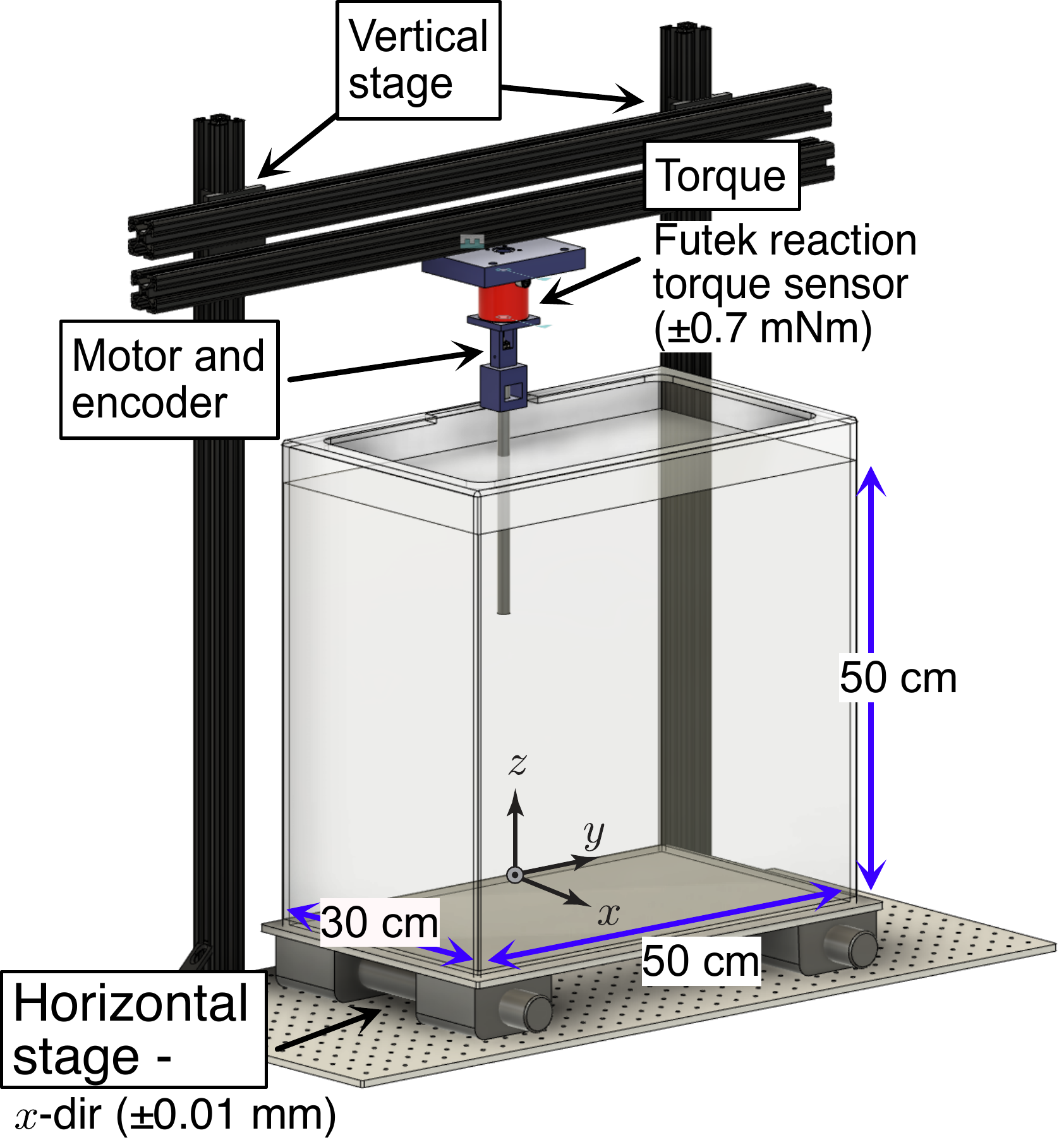}
    \caption{Experimental setup showing the tank, translation stage, torque sensor  and a cylinder positioned for measurement. 
    The motor and magnetic encoder were housed inside a 3D-printed structure that was mounted to the active side of the torque sensor.
    Signal wires were run through the center of the torque sensor for motor control and data acquisition.}
    \label{fig:expt_setup}
\end{figure}

To measure the dependence of torque on boundary distance, we secured the tank onto a horizontal stage that allowed for motion in the $x$-direction, as shown in Fig. \ref{fig:expt_setup}. 
The motion of the stage was controlled by a linear guide with a worm gear screw that advanced the stage 0.3 mm per revolution. 
The screw was turned using a computer-controlled NEMA 23 stepper motor with a resolution of 400 steps/rev. This gave better than 100 $\mu$m precision in controlling the boundary distance, which was necessary: the step size near the boundary was as small as 0.5 mm.
            
Torque measurements were made for both cylinders and helices using similar methods. 
The objects were held in a rigid shaft adapter and then lowered until centered in the tank using a vertical translation stage built from 80-20\textsuperscript{\textregistered} extruded aluminum.   
 
At the beginning of each data set, we first adjusted the vertical tilt of the object until it was parallel to the boundary. 
Next, we manually adjusted the horizontal stage so that the cylinder or helix touched the front vertical boundary of the tank. 
We used total internal reflection to form an image of the object that could be used as a reference to find where the edge of the object just made contact with the boundary, which occurs when the image appears to touch the object.  
                       
The torque was measured using a  FUTEK TFF400, 10 in-oz, Reaction Torque Sensor. 
The cylinder and helices were driven by a variable speed DC motor with a magnetic encoder (Pololu 298:1 Micro Metal Gear Motor with Magnetic Encoder) and housed inside of a 3D-printed enclosure that included sleeve bearings to minimize frictional torque. 
The power and signal wires were fed through a 6.32 mm opening at the center of the torque sensor.
The wires were then fixed to the outside structure so that they did not create a torque when measurements were taken.  
The encoder output was read by the counter input on a National Instruments USB6211 M series multifunction DAQ.  The torque signal was amplified using an amplifier/driver (Omega DP25B-E-A 1/8 DIN Process Meter and Controller) and its output fed into the same National Instruments data acquisition board's analog to digital input with a resolution of 250 thousand samples per second, which is much faster than any time scales in the experiment. 
            
\begin{figure}
    \centering
    \includegraphics[width=\textwidth]{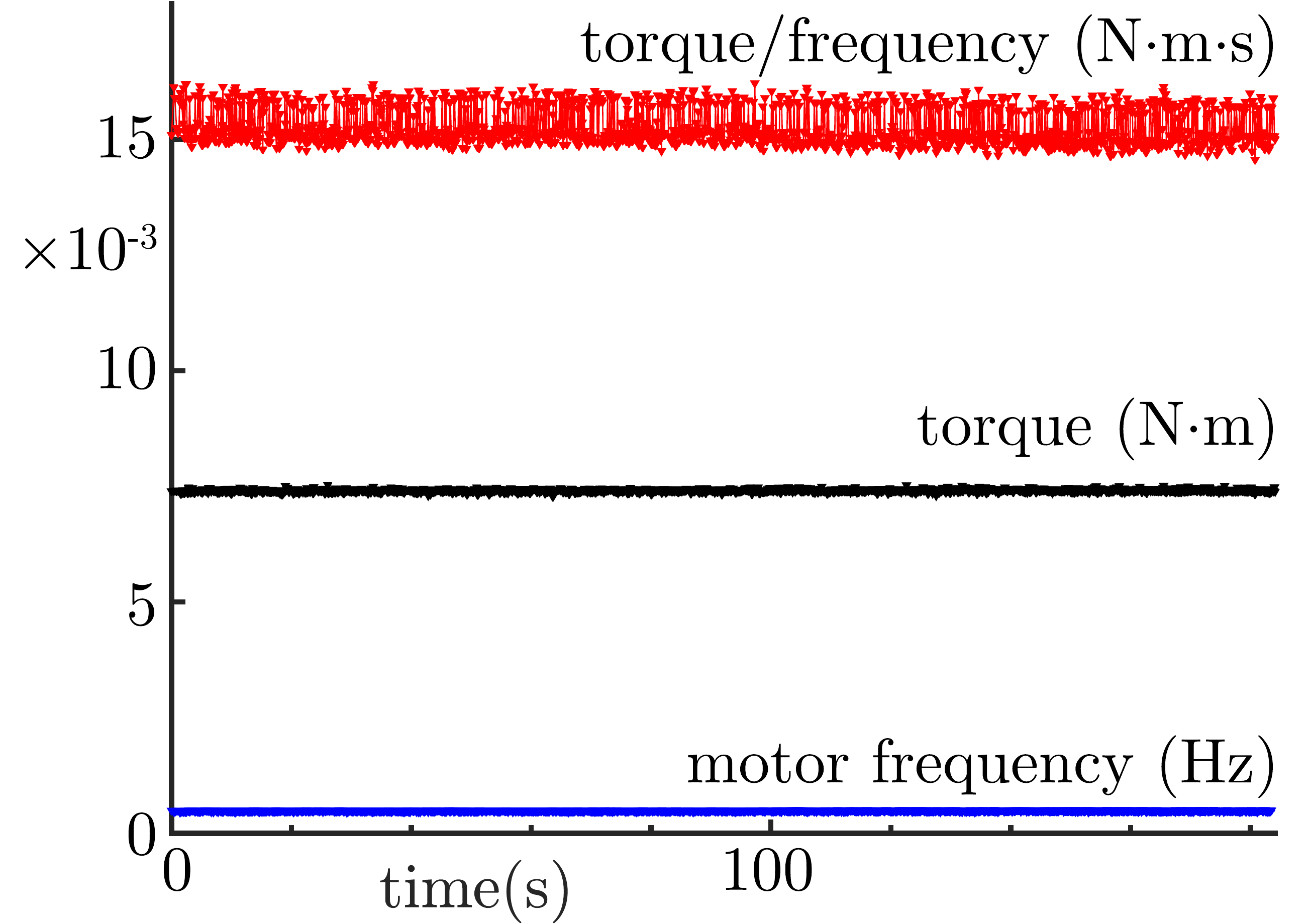}
    \caption{Example of data signals from the experimental torque measurements. 
    The frequency and torque data were read by the DAQ and the torque per unit frequency was calculated in real time and was smoothed to remove outliers as described in the text.
    }
    \label{fig:exp_data}
\end{figure}

Data were taken with the DC motor rotating at varying speeds and with the objects located at a distance from the boundary set by the horizontal stage.
The torque and motor frequency were simultaneously recorded using MATLAB to acquire and plot them.  
We used MATLAB and a motor controller (ARDUINO MEGA 2560 with an ADAFRUIT Motor Shield v.2) to control the motor.
However, the motor rotation varied depending on the axial load, so we divided the signal from the torque sensor by the frequency data from the counter input to get the torque per unit frequency at each boundary distance, see Fig. \ref{fig:exp_data}.
A MATLAB data acquisition GUI included the temperature and distance values, ensuring that the acquisition parameters were stored with the raw data.             
            
Data were taken for approximately 60 rotation periods for both CW and CCW rotation at each boundary location.
The frequency signal occasionally showed large spikes that affected the average torque-per-frequency value because the torque signal did not show a corresponding jump. 
We considered this to be the result of the encoder miscounting the rotation rate or the counter input in the DAQ misreading the signal from the encoder.
We used MATLAB's outliers function to remove such frequency spikes that were more than nine median absolute deviations from the median calculated in a moving window ten data points wide, and replaced them with the average of the adjoining data points. 
The number of outliers was less than 1\% of the data points, so this frequency smoothing should not have biased the averaging significantly.
            
The difference between mean CW and CCW rotation values, which should have been the same, was used to establish the uncertainty in the experimental measurements. 
An analysis script read the geometric parameters and data files for a given set of measurements (cylinder or helix) and plotted the data versus boundary distance. 
We scaled the torque using a unit of $[\mu \Omega r^2\ell]$ (cylinder) or $[\mu \Omega R^2L]$ (helix), where $\mu$ is the fluid viscosity, $\Omega$ is the angular speed, $r$ is cylindrical radius, $\ell$ is the cylindrical length, $R$ is the helical radius, and $L$ is the helical axial length. Plots of the dimensionless torque for cylinders and helices are shown in Fig. \ref{fig:cyl_torque} and Fig. \ref{fig:helix_torque}. 
Using these units for the torque allowed for easy comparison  between experiments, numerical simulations, and theory.             

\subsection{Summary of algorithms and data analysis}

Two separate sets of simulations are  presented in this paper. For those with a helix model or a bacterium model, the results were averaged over 16 evenly spaced phases as described in Eq. \ref{Eq_flag} of the flagellar centerline.

(i) The goal of the first set of simulations was to calibrate the MRS and MIRS methods by finding the optimal factors ($\gamma_c$ for a cylindrical cell body and $\gamma_f$ for a helical flagellum) and the optimal regularization parameters ($\epsilon_c$ and $\epsilon_f$), as reported in Table \ref{Table_bac_model}. 
Eq. \ref{Eq_MRS_velocity} was used to solve for the force $\mathbf{f}_k$ at each discretized point $\mathbf{x}_k$ in a free space, whereas Eq. \ref{Eq_MIRS_velocity} was used for simulations near a plane wall. 
The resulting net torque of each rotating structure was then compared with the results from theory for a cylinder or from experiments for a helix, as described in 
Sec. \ref{sec:blob_size}.

(ii) The goal of the second set of simulations was to assess the motility performance of the force-free and torque-free bacterium models with boundary effects incorporated. 

\textit{Step 1}: Eq. \ref{Eq_MRS_force_torque_free} was used with $S_\epsilon$ (for simulations in a free space) or with $S^*_\epsilon$ (for simulations with a plane wall). 
Different combinations of the cell body size, flagellar wavelength, and distance to the wall were simulated. 
We used five values for the length $\ell$ and five values for the radius $r$ shown in Table \ref{Table_bac_model}.
These values are within the range of normal \textit{E. coli} \cite{Darnton2007flagparameters}.
We used $18$ wavelengths $\lambda$ that cover a range of biological values ($2.22 \pm 0.2$ $\mu m$) and values that are shorter and longer than the biological values (Table \ref{Table_bac_model} and Fig. \ref{fig:bacterium_models}).
The set of geometric parameters, together with $22$ distance values $d$ measured from the flagellar axis of symmetry to the wall, resulted in $9,900$ simulations. 
From each simulation, we obtained the axial component of the translational velocity $U$, 
the magnitude of the axial-component of the 
hydrodynamic drag
on the cell body $F$, and the magnitude of the axial-component of the 
hydrodynamics
torque on the cell body $\tau$.
For each body geometry (450 total), we performed a simulation in free-space to ensure the convergence of MIRS calculations to MRS calculations as the distance $d\rightarrow\infty$. 

\textit{Step 2}: The torque value $\tau$ was output from each simulation in \textit{Step 1} with the motor frequency set to $154\,\text{Hz}$. 
That torque-frequency pair was then used to determine the load line and its intersection with the torque-speed, as discussed in Sec. \ref{sec.torque-speed} and shown in Fig. \ref{fig:torque_speed_curve}. 
Each motor frequency $\Omega_m / 2 \pi$
on the torque-speed curve was given as some multiple $q$ of $154\,\text{Hz}$.
The simulation outputs were scaled by $q$, since they were all linear with motor frequency; i.e., $\left(U, F, \tau\right)\rightarrow q\left(U, F, \tau\right)$. 
These scaled quantities were then used calculate the performance measures. 
Results are presented in Sec. \ref{sec:results_speed} and Sec. \ref{sec:results_energycost}. 
 

\section{Results}\label{sec:results}

\subsection{Verifying the numerical model and determining the optimal regularization parameters}\label{sec:blob_size}

 When using MRS or MIRS, the choice of the regularization parameter for a given discretization (cylinder) or filament radius (helix) of the immersed structure has generally been made without precise connection to real-world experiments, because there are large uncertainties in biological and other small-scale measurements. 
We therefore used theory, as described below, and dynamically similar experiments, as described in Sec. \ref{sec:expts}, to determine the optimal regularization parameters for the two geometries used in our bacterial model: a cylinder and a helix.

\subsubsection{Finding the optimal regularization parameter for a rotating cylinder}\label{sec:cyl_blob}
Jeffrey and Onishi (1981) derived a theory for the torque per length on an infinite cylinder rotating near an infinite plane wall \cite{JeffreyOnishi1981cyltorque} that was used previously to calibrate numerical simulations of helical flagella \cite{das2018computing}. 
The torque per unit length $\sigma$ on an infinite cylinder is given as
\begin{equation}\label{eq:cyl_theory}
    \sigma =4\pi \mu \Omega r^2 \frac{d}{(d^2-r^2)^{1/2}}
\end{equation}
where $\mu$ is the dynamic viscosity of the fluid, $\Omega$ is the angular rotation speed, $r$ is the cylindrical radius, and $d$ is the distance from the axis of symmetry to the plane wall.

We used this theoretical value as a common reference point between the experiments and simulations to establish optimal computational parameters, but note that
this theory has not been experimentally tested outside of the present work. 
We assumed Eq. \ref{eq:cyl_theory} is valid for our experiments and simulations, though this assumption as applied to experiments ignored the finite size of the tank. To control for end effects in the experiments, we measured the torque with only the first 3 cm inserted into the fluid and with the full cylinder inserted at the same boundary locations. We subtracted the torque found for the short section from the torque found for the full insertion of the cylinder. In simulations, we controlled for finite-length effects by measuring the torque on a middle subsection of the simulated cylinder, as discussed below.

\begin{figure}[ht]
\includegraphics[width=\textwidth]{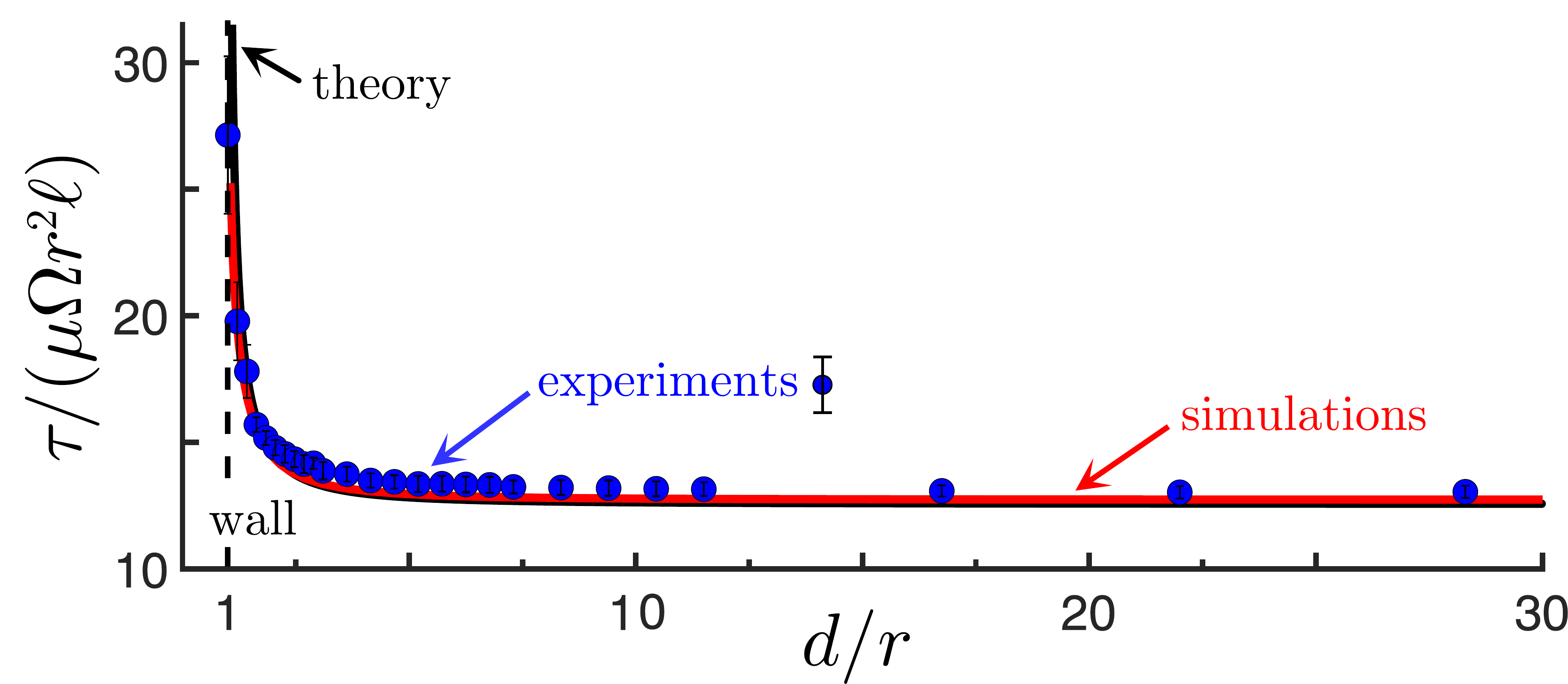}
\caption{Dimensionless torque, $\tau/(\mu\Omega r^2\ell)$, for a cylinder versus scaled boundary distance ($d/r$), where $\mu$ is the dynamic viscosity, and $\ell$ is the length of the cylinder. 
The boundary distance is scaled by the cylindrical radius $r$ as measured to the centerline of the cylinder: theory by Jeffrey and Onishi \cite{JeffreyOnishi1981cyltorque} (solid black line), optimized MIRS simulations (solid red curve) and dynamically similar experiments (solid blue circles). 
The numerical simulations were optimized by adjusting the discretization factor $\gamma_c$ to minimize the MSE between theory and simulation (the minimum MSE is 0.36\%). 
The MSE between experiments and theory was large near the boundary because the theory goes to infinity at $d/r =1$.
Outside of the near-boundary region ($d/r\ge 2$) the MSE is less than 1\%.}\label{fig:cyl_torque}
\end{figure}

Our experimental data are shown in Fig. \ref{fig:cyl_torque}, with the torque made dimensionless using the quantity $\mu\Omega r^2\ell$, where $\mu$ is the fluid viscosity, $\Omega$ is the rotation rate, $r$ is the cylindrical radius, and $\ell$ is the cylindrical length. 
The mean squared error (MSE) between experiments and theory is MSE $\leq 6\%$ when calculated for the boundary distances where $d/r>1.1$ (i.e. the distance from the boundary to the edge of the flagellum is $\geq 1$ mm). 
The theory asymptotically approaches infinity as the boundary distance approaches $d/r = 1$, which skewed the MSE unrealistically. 
For the data where $d/r\ge 2$, the mean squared error is less than 1\%. 

In numerical simulations of the cylinder, the computed torque value depended on both the discretization and regularization parameter. 
Having found good correspondence with the experiments, we used Eq. (\ref{eq:cyl_theory}) to find an optimal regularization parameter for a given discretization of the cylinder (see Table \ref{Table_bac_model}: cylinder part). 
The discretization size of the cylindrical model $ds_c$ was varied among $0.192$ $\mu$m, $0.144$ $\mu$m, and $0.096$ $\mu$m. 
For each $ds_c$, an optimal discretization factor $\gamma_c$ was found by minimizing the MSE between the numerical simulations and the theoretical values using the computed torque in the middle two-thirds of the cylinder to avoid end effects. The optimal factor was found to be $\gamma_c=$ 6.4 for all the discretization sizes. 
We used the finest discretization size for our model bacterium as reported in Table \ref{Table_bac_model} since it returned the smallest MSE value of 0.36\%. 

\subsubsection{Finding the optimal regularization parameter for a rotating helix far from a boundary}\label{sec:helix_blob}
Simulated helical torque values also depend on the discretization and regularization parameter, but there is no 
theory for a helix to provide a reference.
Other researchers have determined the regularization parameter using complementary numerical simulations, but the reference simulations also have free parameters that may have affected their results \cite{martindale2016blobsize}.

\begin{figure}[ht]
\includegraphics[width=\textwidth]{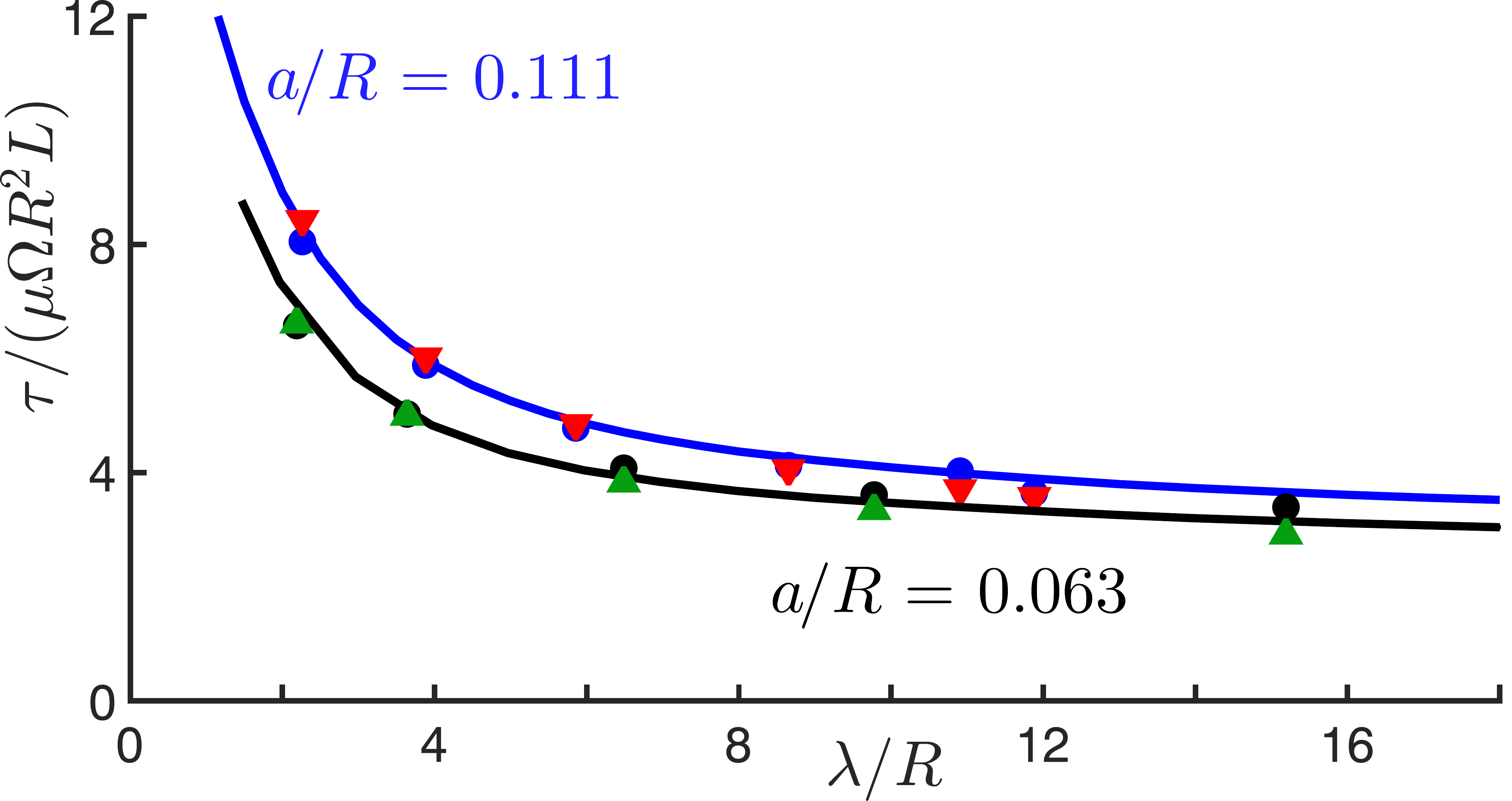}
\caption{Dimensionless torque ($\tau/(\mu\Omega R^2L)$ for different flagellar wavelengths ($\lambda/R$), where $\mu$ is the dynamic viscosity, $\Omega$ is the angular speed, $R$ is the helical radius and $\lambda$ is the helical wavelength. 
Experimental values are solid black circles and solid blue circles; our MRS simulations with a centerline distribution of regularized Stokeslets are the solid red and solid green triangles, and MRS simulations computed with a surface discretization of the helices using the code provided by Rodenborn et al. (2013) \cite{rodenborn2013propulsion} are the  blue and black curves. }
\label{fig:free_torque}
\end{figure}

Thus, we used dynamically similar experiments, as described in Sec. \ref{sec:expts}, to determine the optimal filament factor, $\gamma_f=2.139$, for a helix filament radius $a/R = 0.111$. Torque were measured for the six helical wavelengths given in Table \ref{tab:flagella} when the helix was far from the boundary. The optimal filament factor $\gamma_f=2.139$ was found by the following steps (i) varying $\epsilon_f$ for each helix until the percent difference between the experiment and simulation was under 5\%; and (ii) averaging the $\epsilon_f$ values found in Step (i). In these simulations, the regularization parameter and discretization size are both equal to $\gamma_f a$.
The results are shown in Fig. \ref{fig:free_torque}, with the torque values  non-dimensionalized by the value $\mu\Omega R^2L,$ where $\mu$ is the fluid viscosity, $\Omega$ is the rotation rate, $R$ is the helical radius, and $L$ is the axial length. The optimized simulations returned an average percent difference of $2.4\pm 1.7$\% compared to the experimental values. 

We checked whether helices with different filament radii could be accurately simulated using our optimized $\gamma_f$ to scale the  regularization parameter, i.e. ($\epsilon_f = \gamma_f a$) to account for relative size of the filament, as is commonly done \cite{Jabbarzadeh2020inextensible, Olson2011sperm, Nguyen2019choano, Buchmann2018pairsofhelices, Bouzarth2011ModelingSB}. We computed torque values that matched the experimental values given in Rodenborn et al. (2013), which used a filament radius $a/R=0.063$. The results are also presented in Fig. \ref{fig:free_torque}. The percent difference between our MRS simulations and their data is 2.5$\pm$1.3\%. 

Martindale et al. (2016) \cite{martindale2016blobsize} used an MRS with a surface discretization of the flagellum to calibrate their simulation parameters, whereas our MRS used a string of regularized Stokeslets along the helical centerline to reduce the computational cost in the MIRS calculations. As a final test, we used the freely available and calibrated code for the MRS with surface discretization from Rodenborn et al. to compute torque values for our $a/R = 0.111$ data and for their $a/R=0.063$ data. Fig. \ref{fig:free_torque} shows the torque comparison of their surface discretized MRS (solid curves), our centerline distribution MRS (triangles), and the experiments (circles). The percent difference between their MRS and the experiments is $3.6\pm3.4$\%. The percent difference between our MRS and theirs is 1.8$\pm$3.7\%. Thus, within the range we tested our MRS with a centerline distribution, the optimal filament factor $\gamma_f$ worked very well for another filament radius and other helical wavelengths when compared to both the experiments and the surfaced discretized MRS in Rodenborn et al. (2013) \cite{rodenborn2013propulsion} for torques far from the boundary.

\subsubsection{Torque on rotating helices near a boundary}
\begin{figure}[ht]
\includegraphics[width=\textwidth]{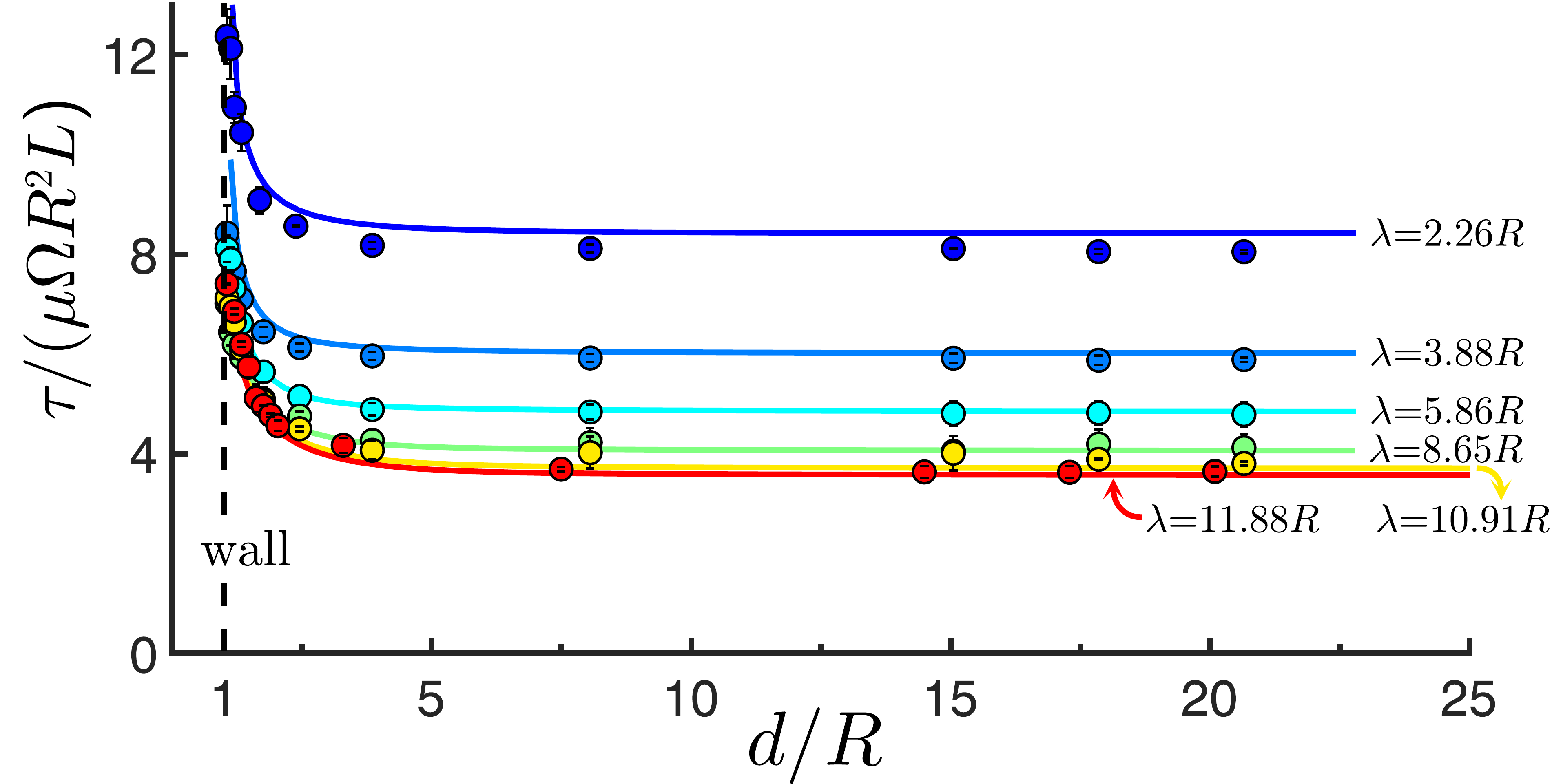}
\caption{Dimensionless torque $\tau$ for different helical wavelengths ($\lambda/R$) versus boundary distance ($d/R$) scaled by the helical radius $R$. 
The optimized MIRS simulations are the solid curves and the experimental values are solid circles with vertical error bars. 
The data also show good agreement for the far from boundary value at $d/R\approx 20$ (see Fig. \ref{fig:free_torque}). The data show that once the far from boundary distance was properly calibrated, the MIRS worked very well to represent the effects of the boundary.}\label{fig:helix_torque}
\end{figure}

To determine how boundaries affect bacterial motility, we used our optimized value for $\gamma_f$ in our MIRS simulations to compute the torque as a function of boundary distance, as shown in Fig. \ref{fig:helix_torque}. The computed torque values and measured torque values also show excellent agreement at most boundary distances, except for the shortest wavelength $\lambda/R=2.26$. 
We note that this helix had the largest variation in wavelength, as reported in Table \ref{tab:flagella}. Furthermore, the torque for short wavelengths is more sensitive to variation in wavelength as compared to variation at longer wavelengths, which likely explains the difference between simulation and experiment for this geometry, whereas for the other wavelengths the simulated values are generally within the uncertainty in the experiments for all boundary distances. 

\subsection{Speed Measurements to Assess Performance} \label{sec:results_speed}
The motion of bacteria through their environment enables them to find nutrients.
Indeed, it has been suggested that the purpose of bacterial motility is primarily to perform chemotaxis \cite{purcell1977life}. 
Living in a microscopic environment where thermal effects are significant, bacteria must be able to sample chemical concentrations faster than diffusion causes those concentrations to change \cite{purcell1977life, schuech2019motile}, so moving faster may confer a survival advantage.

The low speed operating regime of the bacterial motor (below 175 Hz) is thermodynamically more efficient than the high speed regime. 
A simple model gives the fraction of energy lost to friction in the motor as
$(\tau_0 - \tau)/\tau_0$, where $\tau_0$ is the stall torque and $\tau$ is the operating torque at a given frequency \cite{li2006low}.
In the low speed regime $\tau\ge0.92\tau_0$, so that the power output of the motor is greater than 92\% of the power input.
However, the low speed regime may be less operationally reliable for motility; the flatness of the torque-speed curve implies that small increases in load correspond to large decreases in motor rotation rate, so the bacterium risks stalling and may be unable to restart its motor. 
Using our simulations, we determined the swimming speed and motor rotation rate for different bacterial geometries at different distances to a solid boundary and
assessed. 
the performance of bacterial geometries typically associated with swimming.

\subsubsection{Optimal flagellar wavelength}

\begin{figure}[ht]
\includegraphics[width=.6\textwidth]{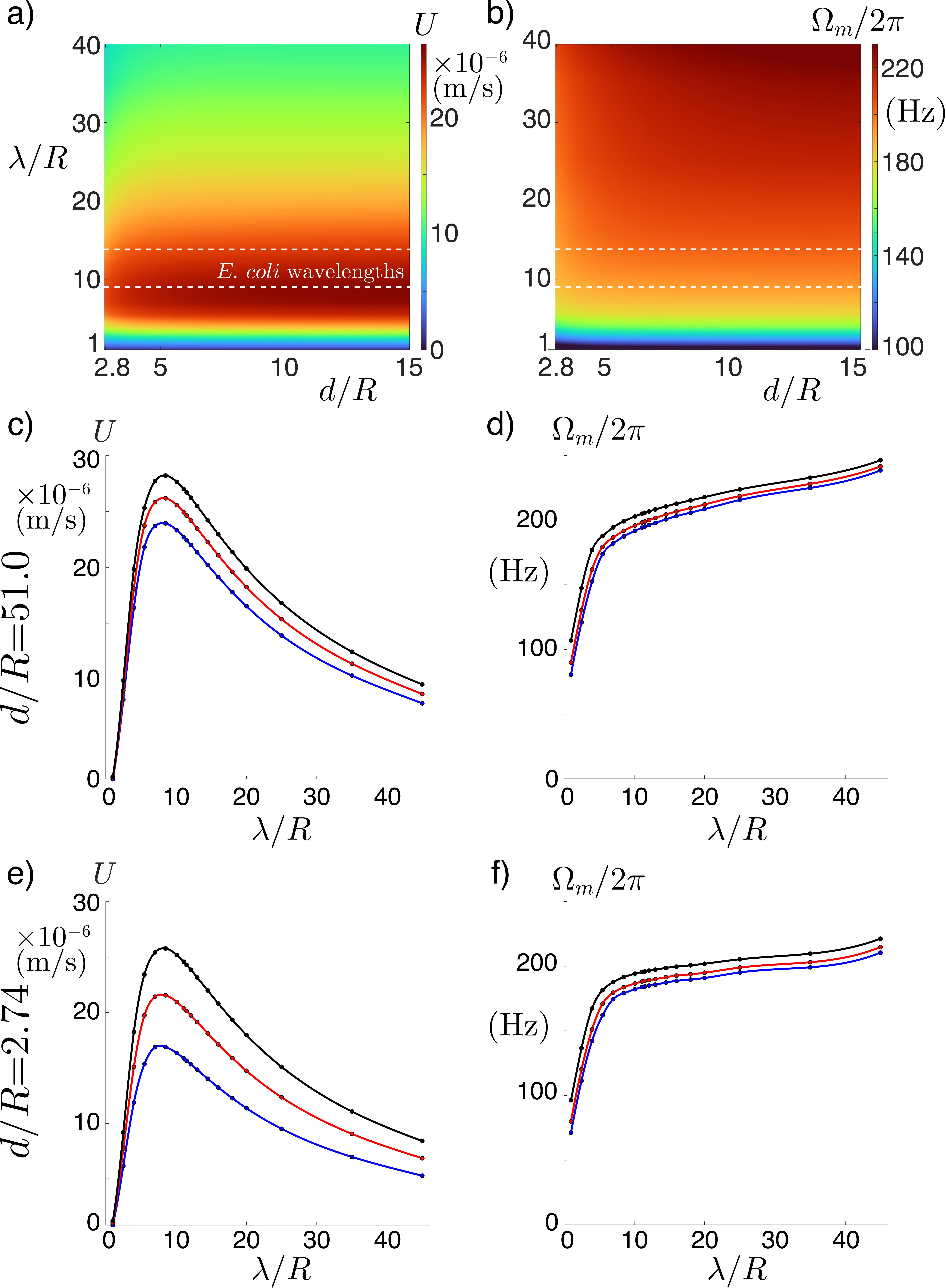}
\caption{Swimming speed and motor
frequency
for different flagellar wavelengths at different boundary distances. 
Panels a and b show heat maps of free swimming speed $U$ and motor frequency $\Omega_m/2\pi$ with axes flagellar wavelength ($\lambda/R$) versus boundary distance ($d/R$), where $R$ is the helical radius.  Typical \textit{E. coli} wavelengths are indicated with the dashed white lines, which shows this range is near to the peak in swimming speed. Panels c-e show line plots of speed and motor frequency across different body sizes far from ($d/R=51.0$) and near ($d/R=2.74$) the boundary. The solid circles are the simulation data points and the solid curves are spline fits to the data. The three curves show the maximum  (black), the median (red) and the minimum (blue) among all cell bodies simulated. Panels c and e show that the peak swimming speed $\lambda/R\approx8$, which is close to the range of \textit{E. coli} wavelengths and the peak has a long ``tail'' as wavelength increases. Panels d and f show increasing motor frequency with increasing wavelength. The trend reflects the plot of the torque-speed curve in Fig. \ref{fig:torque_speed_curve}.
}
\label{fig:UandOmega}
\end{figure}

We first consider the 
effect of different flagellar wavelengths
on swimming speed and motor rotation rate, as shown
in Figs. \ref{fig:UandOmega}a and b. 
Swimming speed and motor rotation rate are shown as heat maps for different flagellar wavelengths at different distances to the boundary. The 
heat map 
shows the median values computed among all 25 bacterial body geometries we investigated (Table \ref{Table_bac_model}). 

The maximum of all the median swimming speed values is about $26\,\mu\text{m}s^{-1}$, and it occurs far from the boundary for a wavelength near $8R$. 
For long and short flagellar wavelengths, the swimming speed at all distances is much lower than the maximum. 
Long wavelengths yield about $10\,\mu\text{m}s^{-1}$, whereas very short wavelengths give values closer to $1\,\mu\text{m}s^{-1}$.
For the flagellar wavelength of $\lambda/R =11.1$ that is typical for \textit{E. coli}, the swimming speed is about $25\,\mu\text{m}s^{-1}$ far from the boundary, whereas it drops to about $20\,\mu\text{m}s^{-1}$ very near the boundary. 

Interestingly, the flagellar wavelengths that correspond to swimming speeds near the maximum in Fig. \ref{fig:UandOmega}a also correspond to motor rotation rates in the low end of the high speed regime in the torque-speed curve, so that the motion is both thermodynamically efficient and operationally reliable. 
A wavelength of $\lambda/R = 8$ gives 190 Hz and 183 Hz far from and near to the wall, respectively, which correspond to mechanical energy outputs of about 84\% and 88\%. Short and long wavelengths result in a weaker performance, but for different reasons: short wavelengths operate in the low speed regime and thus are efficient but unreliable, whereas longer wavelengths operate farther into the high speed region and thus are reliable but inefficient.

\subsubsection{Boundary effects}

To illustrate how proximity to the boundary affects swimming speed and motor rotation rate, we show line plots in Figs. \ref{fig:UandOmega}c-f of the speed and rotation rate as functions of the flagellar wavelength both far from and near to the boundary. 
The maximum, median, and minimum values among all bacterial body geometries are shown for each boundary distance.
Comparing Figs.\ref{fig:UandOmega}c and e shows that proximity to the boundary does not appreciably alter the optimal wavelength: it remains near $8R$ for all body geometries both near and far from the boundary. 
However, proximity to the boundary does increase the difference in the swimming speed among different bodies at a given wavelength. 

Far from the boundary, the difference between the maximum and minimum swimming speeds for the optimal flagellar wavelength is 14\% of the maximum value of $28\,\mu\text{m}s^{-1}$; near the boundary, the difference is 34\% of the maximum value of $26\,\mu\text{m}s^{-1}$. 
Figs. \ref{fig:UandOmega}d and f show
the motor rotation rate is less sensitive to the body geometry and proximity to the surface than the swimming speed. 
Far from the boundary, the difference between the maximum and minimum rotation rates for the optimal flagellar wavelength $8R$ is 6\% of the maximum value $198\,\text{Hz}$; near the boundary, the difference is 6\% of the maximum value of $190\,\text{Hz}$. 

\begin{figure}
\includegraphics[width=\textwidth]{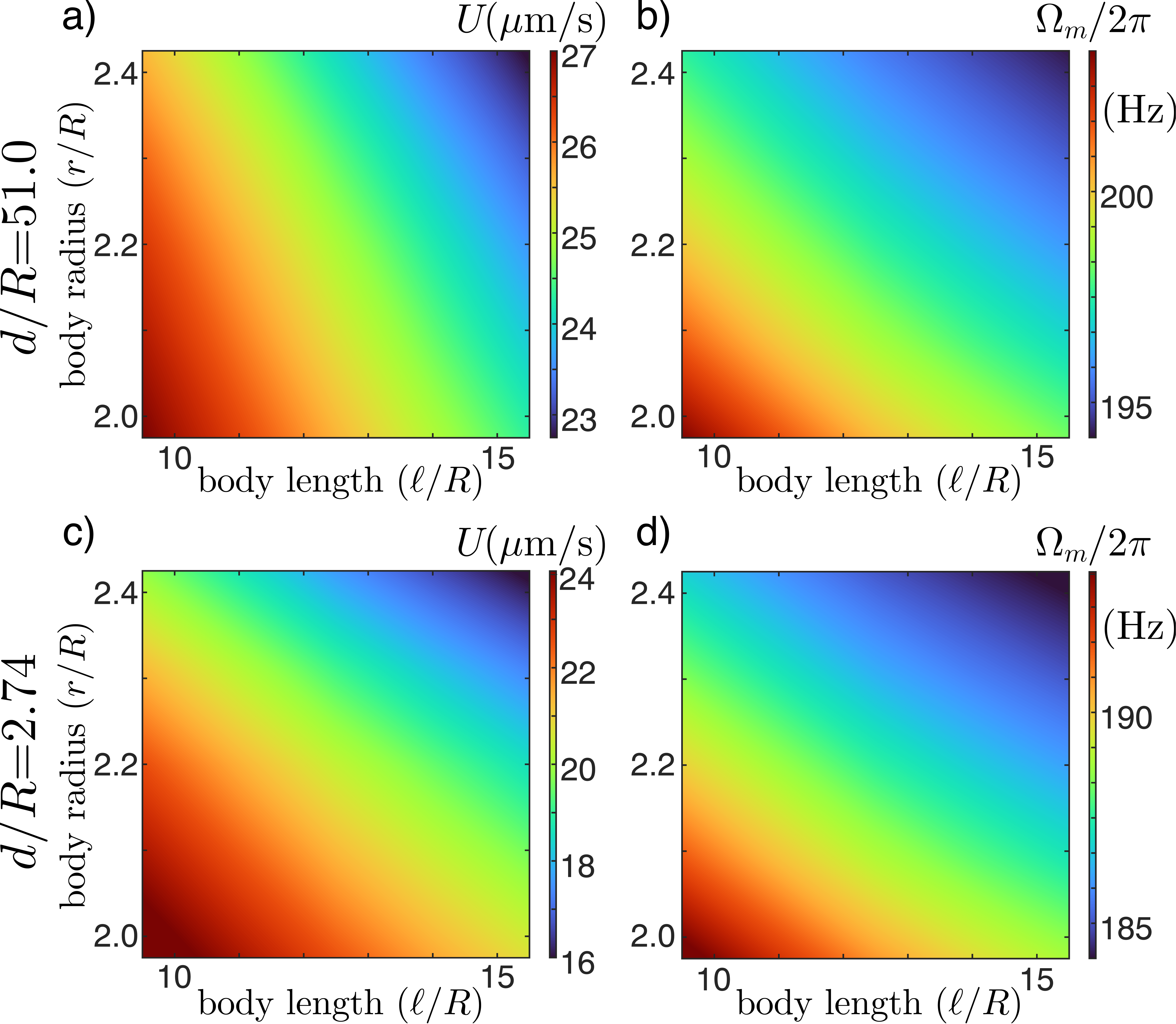}   
\caption{Free-swimming speed $U$ and motor frequency $\Omega_m/2\pi$ shown as heat maps with axes cylindrical radius ($r/R$) versus body length ($\ell/R$). 
The data are for a fixed flagellum wavelength, $\lambda/R =11.1$, where $R$ is the helical radius. The top row a and b is far from the boundary $d/R=51.0$ data where boundary effects are minimal, and the bottom row c and d are data close to the boundary $d/R=2.74$. The swimming speed data in a and c show that short thin
bodies result in higher swimming speed both near and far from the boundary, though near the boundary the swimming speed is lower for a given body geometry. Therefore the swimming speed measure predicts short thin bodies far from the surface result in a better motility performance. The motor frequency data in b) and d) show long thick bodies result in a slower motor frequency near and far from the boundary, though far from the boundary the motor frequency is higher. Therefore, the motor frequency measure predicts long thick bodies near the surface result in better motility performance.
}
\label{fig:UandOmegaBodyMaps}
\end{figure}

To further probe the effect of the cell body geometry on swimming speed and motor rotation rate, we show heat maps of the speed and rotation rate fixed at the typical \textit{E. coli} wavelength $\lambda/R = 11.1$ as functions of the length and radius of the cylindrical cell body.
Figs. \ref{fig:UandOmegaBodyMaps}a-d show the results. The translational speed is optimized for short thin cell bodies (lower left-hand corner of Figs. \ref{fig:UandOmegaBodyMaps}a and c both near and far from the surface.
Conversely, the slowest motor rotation rates (though all higher than 175 Hz), and therefore the most thermodynamically efficient, occur for long thick cell bodies (upper right-hand corners of Figs. \ref{fig:UandOmegaBodyMaps}b and d.
Taken together, these two results suggest that balancing the need of a bacterium to move quickly with its need to be thermodynamically efficient would yield a cell body geometry somewhere between long, thick cell bodies and 
short, thin cell bodies. 
Interestingly,
the center point of the heat maps shown in Fig. \ref{fig:UandOmegaBodyMaps} corresponds to the mean size of the \textit{E. coli} cell body. 

\subsection{Energy Cost Measures to Assess Performance} \label{sec:results_energycost}
The energy cost required to 
move
is another way to assess the performance of the bacterial motility system.
Here we present simulation results of three different energy cost measures.
The first measure we consider is what we term the Purcell inefficiency $\mathcal{E}_{Purcell}^{-1}$ given by,

\begin{equation}
    \mathcal{E}_{Purcell}^{-1} = \frac{\tau\Omega_m}{FU},
    \label{eq:PurcellInefficiency}
\end{equation}
where $\tau$ is the motor torque (or the torque on the cell body or the flagellum), $\Omega_m$ is the motor rotation rate, $F$ is the drag force on the cell body 
(or on the flagellum), and $U$ is the swimming speed of the bacterium.


Thus, the Purcell inefficiency measures the mechanical energy ($T\Omega_m$ ) required to swim at speed $U$ relative to the least amount of energy ($FU$) needed to translate the cell body at speed $U$. 
The Purcell inefficiency is useful because, under certain simplifying assumptions  \cite{purcell1997efficiency}, it can be expressed as a function of the geometry of the cell body and the flagellum alone. 
The difficulty with this measures
is that it does not depend on the rotation rate of the motor because
all four quantities appearing in Eq. \ref{eq:PurcellInefficiency} scale with the motor frequency (see Eq. \ref{Eq_MRS_force_torque_free}). Therefore, the Purcell inefficiency cannot assess how swimming performance depends on the torque-speed characteristics of the motor, and thus omits an important element of the bacterial motility system that is subject to selective forces.

The second measure is the energy cost $E$ to travel a unit distance $d$ given by 
\begin{equation}
    \frac{E}{d} = \frac{\tau\Omega_m}{U}.
\end{equation}
Several authors \cite{li2006low, li2017flagellar} have considered the distance traveled per energy output by the motor, which is the inverse of the measure we consider here. The merit of the energy cost per distance measure is that it expresses the amount of energy used by the bacterium to perform a biologically relevant task; namely, to swim one unit distance. Another advantage is that it
depends on the motor rotation rate and thus can probe the effect of the torque-speed characteristics of the motor. However, it does not account for the size of the bacterium, and thus does not measure the energy cost relative to the overall metabolic budget of the organism.

To account for the metabolic energy cost required to swim a unit distance, we introduce a third measure,
\begin{equation}
    \frac{(E/m)}{d} = \frac{\tau\Omega_m}{mU}.
\end{equation}
The mass $m$ associated with each bacterial model is $m = 1.1\times10^{-15} \left(\pi r^2l\right)\,\text{kg}$, where $r$ is the body radius and $\ell$ is the body length, both measured in $\mu\text{m}$. Though this energy cost measure has not been considered in the literature, it was suggested earlier by Purcell \cite{purcell1977life}. 

\subsubsection{Optimal wavelength}
We first consider the optimal flagellar wavelength predicted by the three energy cost measures, as shown in Figs. \ref{fig:EnergyCostsLambda}.
The top row a-c shows heat maps of the three energy cost measures as functions of flagellar wavelength and boundary distance, which correspond to the median values computed for all body geometries listed in Table \ref{Table_bac_model}. All three measures give an optimal wavelength near $\lambda/R = 8$ (where each energy cost measure is minimal). However, the three measures differ in other ways. 
The Purcell inefficiency predicts that swimming near the boundary is less inefficient than swimming far from the boundary, whereas the opposite is true for the energy per distance and metabolic cost measures. At a wavelength of $8R$, the minimum Purcell inefficiency value is about 84 (or $1/84 = 1.2\%$ if calculated as Purcell efficiency), the minimum energy per distance measure is $5.0\times10^{-11}\,\text{Jm}^{-1}$, and the minimum metabolic energy cost is $3.1\times10^{4} \,\text{J}\text{m}^{-1}\text{kg}^{-1}$. 

\subsubsection{Boundary effects}
\begin{figure}
\includegraphics[width=\textwidth]{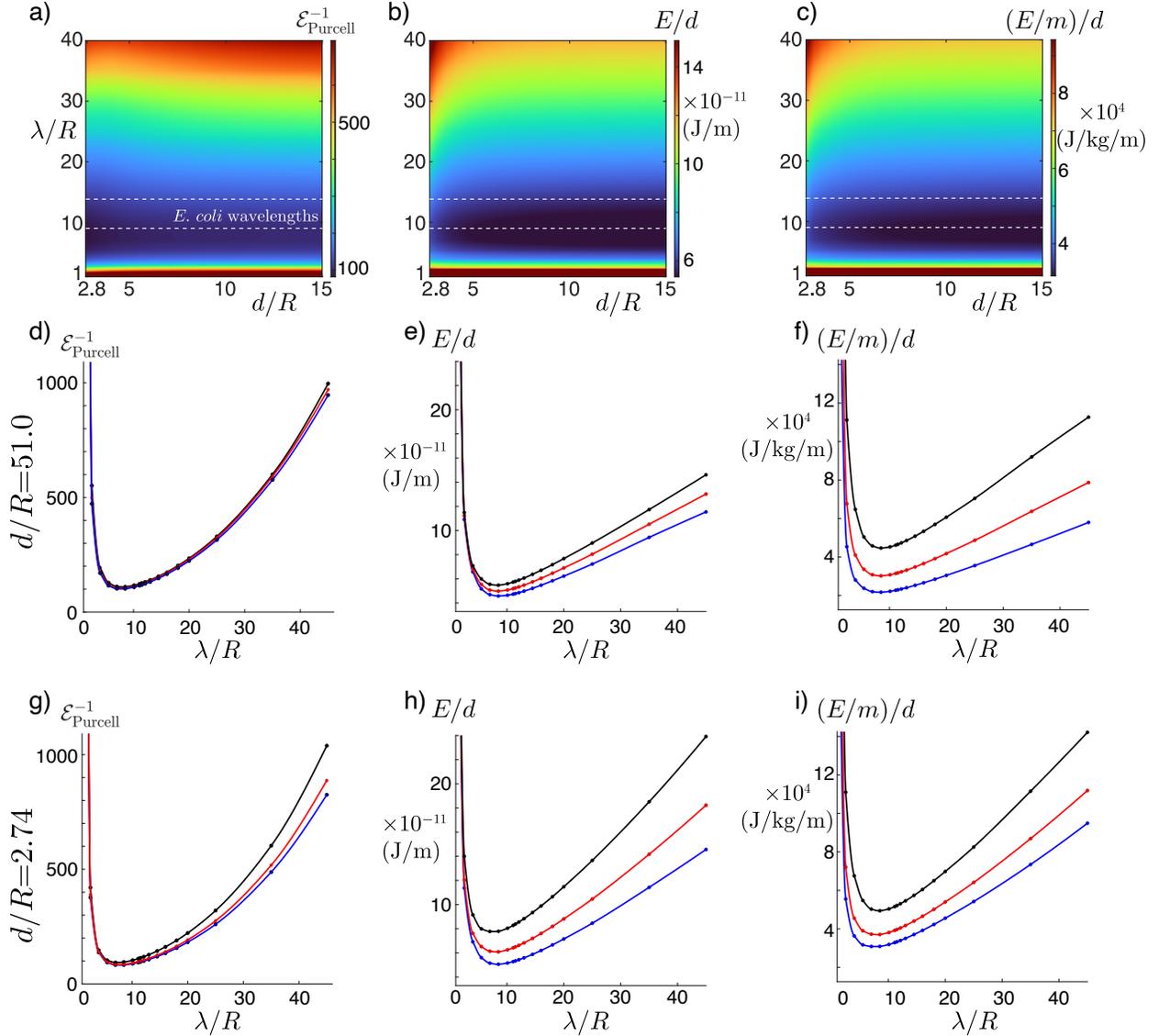}
\caption{Energy cost as a function of wavelength and boundary distance. The top row shows three energy cost measures as a function of helical wavelength $\lambda/R$ and boundary distance $d/R$, where $R$ is the helical radius. Typical \textit{E. coli} wavelengths are indicated with the dashed white lines whose range is close to the optimal wavelength predicted by these energy cost measures. The second and third rows show line plots at distances far from ($d/R =51.0$) and near ($d/R=2.74$) the boundary to assess the wavelength dependence of each measure at those distances. The solid circles are numerical simulations and the solid curves are spline fits to the numerical data. The three curves show the maximum  (black), the median (red) and the minimum (blue) among all cell bodies simulated. All these plots have the optimal flagellar wavelength $\lambda/R \approx 8$.
}
\label{fig:EnergyCostsLambda}
\end{figure}

To evaluate how proximity to the surface affects the predictions of the energy cost measures, we show line plots in Figs. \ref{fig:EnergyCostsLambda} of the measures as functions of flagellar wavelength far from ($d/R = 51$) in d-f and near the boundary ($d/R =2.74$) in g)-i). The maximum, median, and minimum values among all body geometries are shown for each wavelength. 
The Purcell inefficiency is the least sensitive of the measures to changes in the body size. For a wavelength of $\lambda/R = 8$, the difference between the maximum and the minimum is 8\% of the maximum (110 vs 101). Near the boundary, the difference increases to 13\% of the maximum value (94 vs 82).

The energy per distance measure is more sensitive to the body size, and the sensitivity increases near the boundary. For a wavelength of $\lambda/R = 8$, the difference between the maximum and minimum values is 16\% of the maximum value of $5.5\times10^{-11}\,\text{J}\text{m}^{-1}$ far from the boundary.
Near the boundary the difference increases to 35\% of the maximum value of $7.5\times10^{-11}\,\text{J}\text{m}^{-1}$. 
The metabolic energy cost is the measure most sensitive to the body size, though interestingly the sensitivity decreases with proximity to the boundary. 
At a wavelength of $\lambda/R = 8$, the difference between the maximum and minimum value far from the boundary is 51\% of the maximum value of $4.5\times10^{4}
\,\text{J}\text{m}^{-1}\text{kg}^{-1}$. Near the boundary the difference decreases to 38\% of the maximum value of $5.0\times10^{4}\,\text{J}\text{m}^{-1}\text{kg}^{-1}$. 

\begin{figure}
\includegraphics[width=\textwidth]{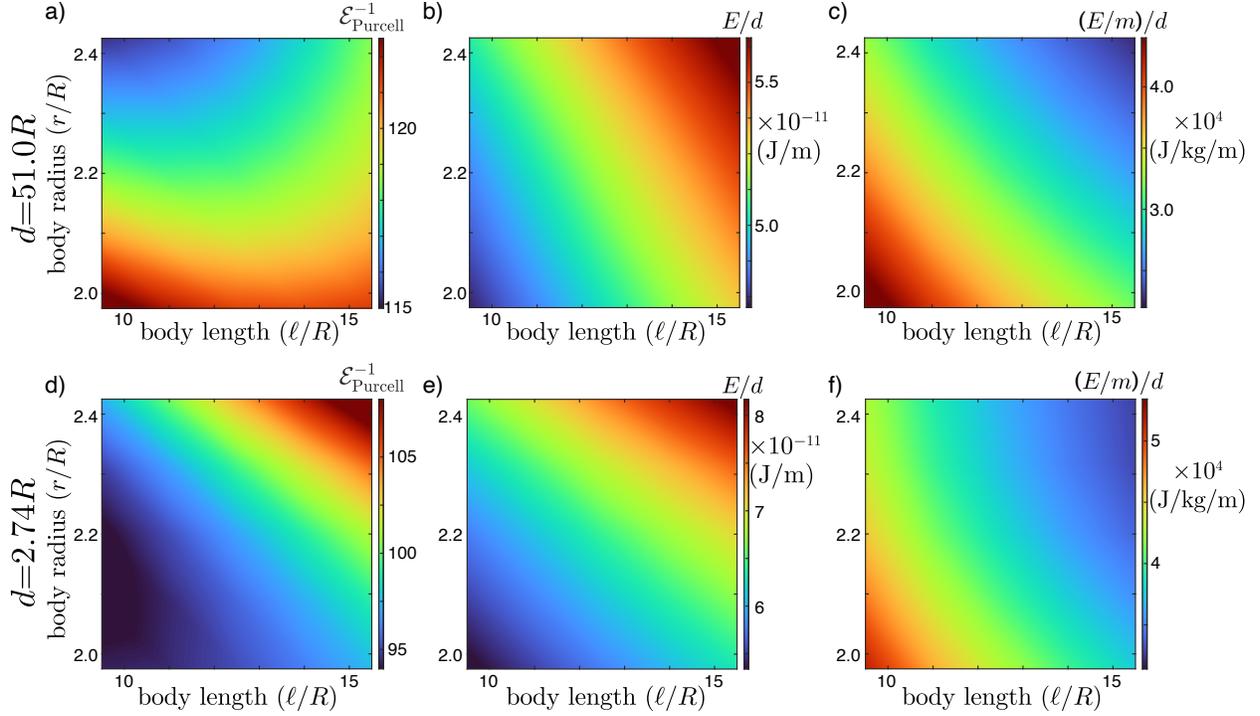}
\caption{Comparison of Purcell inefficiency, energy per distance, and metabolic
energy cost with respect to body geometry at the typical wavelength of \textit{E. coli} ($\lambda/R = 11.1$).
The top row shows results far from the boundary ($d/R=51.0$) and the bottom row shows results near the boundary ($d/R=2.74$). In panels a and d, the Purcell inefficiency shows that short thick bodies are most efficient (i.e., least inefficient) far from the boundary but short thin bodies are most efficient near the boundary. In panels b and e, short thin bodies require the least energy cost per distance both far from and near the boundary. In panels c and f, long thick bodies require the least metabolic energy cost per distance traveled both far from and near the boundary
}
\label{fig:EnergyCostBody}
\end{figure}

Finally, we consider how the energy cost measures depend on body radius and body length at different distances to the boundary.
In Fig. \ref{fig:EnergyCostBody} we show heat maps of the three energy cost measures fixed at the typical \textit{E. coli} wavelength $\lambda/R = 11.1$, as functions of the radius and length. 
The Purcell inefficiency shown in Figs. \ref{fig:EnergyCostBody} gives different optimal body geometries near and far from the boundary: far from the boundary short, thick cylinders (top left corner of Fig. \ref{fig:EnergyCostBody}a) are the least inefficient; near the boundary short thin cylinders (bottom left corner of Fig. \ref{fig:EnergyCostBody}d) are the least inefficient. 
The energy per distance measure gives the same optimal body far from and near to the boundary: the lowest energy per distance cost measure is given by short, thin cylinders (bottom left corners of Figs. \ref{fig:EnergyCostBody}b and e.
The metabolic cost measure gives the same optimal body near and far from the surface, though it is opposite of the optimal body predicted by the energy per distance measure: the lowest metabolic cost measure occurs for cylinders that are long and thick (top right corners of Figs. \ref{fig:EnergyCostBody}c and f.

\section{Discussion}\label{sec:discussion}

In this work we used the method of images for regularized Stokeslets (MIRS) to simulate a motile flagellate bacterium moving near a solid boundary. 
We determined the regularization parameter in the method by conducting dynamically similar macroscopic experiments with rotating cylinders and rotating helices near a solid boundary and comparing the results to equivalent simulations. By varying the regularization parameters, we were able to find optimal values that matched the experimental results within 5\%. Having calibrated MIRS, we simulated various bacterial morphologies to assess their swimming performance. We assessed swimming performance using multiple measures: swimming speed, motor rotation rate, the Purcell inefficiency, energy cost per distance, and metabolic energy cost per distance. As a important and novel addition to our simulations, we incorporated the experimentally measured torque-speed response curve \cite{chen2000torque} by ensuring that the torque and motor rotation rate matched a point on the curve in all our calculated measures. 

Using our MIRS calibration method, we found that the optimal discretization factor for a cylinder is $\gamma_c = 6.4$ for the surface discretizations
we used, which may be used as a reference value for other researchers who simulate rotating cylinders using MRS or MIRS. We also found an optimal filament factor $\gamma_f = 2.139$ when using MRS and MIRS with each helix
modeled as a string of regularized Stokeslets along the helix centerline. 
Selecting an appropriate regularization parameter for 
a center-line discretization of helices MIRS has been considered by other researchers. Martindale et al. (2016) \cite{martindale2016blobsize} benchmarked their center-line discretization of a helix with a surface discretization model. They reported that the optimal filament factor should be in the range $1\leq \gamma_f \leq 3$ to keep the percent difference less than about 10\% in their simulations, which is consistent with our results.

In our work, we calibrated simulations that used a centerline discretization of helices by fitting the regularization parameter directly with experimentally measured values of torque. 
These MIRS computations showed excellent agreement with the experimental torque values at most boundary distances (Fig. \ref{fig:helix_torque}).

In MRS/MIRS, using a centerline distribution for a model helix (or flagellum) with a calibrated regularization parameter is more useful than a surface discretization for several reasons: (i) the computational cost is significantly reduced because the matrix system for the centerline distribution is much smaller than for a surface discretization; (ii) simulations of very short helical wavelengths using a centerline distribution do not encounter discretization issues such as overlapping
cross-sections; 
(iii) in a centerline distribution, the point connecting the cylindrical cell body and the tapered helical flagellum can be considered as the motor location, whereas
the motor location in a surface discretization is hard to define because of the
small gap between the cell body and the flagellum needed
to allow counter-rotation between the cell body and the flagellum.

Interestingly, all five performance measures we computed with our calibrated
model -- swimming speed, motor speed, Purcell inefficiency, energy per distance, and metabolic energy cost -- predict an optimal flagellar wavelength of $\lambda/R \approx 8$, where $R$ is the helical radius of the flagellum. This result agrees with the work of Zhang et al. (2014) \cite{he2014propulsive} who studied the Purcell efficiency of a rotating helix, whereas our model includes a cell body with rotation and translation. Furthermore, this prediction occurs both near and far from the surface and for all body geometries, which suggests that the bacterial wavelengths may be selected independently of body shape or surface proximity.  Therefore, none of the five measures can be distinguished by their predictions of flagellar wavelength, but they are distinguished by their predictions for the optimal body size.

Further analysis showed that the swimming speed is optimal (i.e. fastest) for bodies that are short and thin, both near and far from the surface. The structure of the torque-speed curve imposes two competing conditions that need to be balanced to achieve optimality: at low speeds the torque-speed curve is flat and therefore thermodynamically efficient, but in that regime small increases in applied load result in large decreases in motor rotation rate that could cause the motor to stall. We therefore suggest that the optimal speed is higher than the 175 Hz knee speed (see Fig. \ref{fig:torque_speed_curve}) so that the motor operates in the reliable regime, but not much higher so that it remains thermodynamically efficient. The lowest motor speed that is still above the knee speed for typical bacterial wavelengths occurs for long and thick bacterial bodies, both near and far from the surface. It is tempting to suggest that balancing the short, thin bodies needed for optimal speed and long, thick bodies needed for optimal motor operation yields the average body size. However, we do not infer too much from this result because we do not have a principled way of performing the balancing needed to draw a definitive conclusion.

The three energy cost measures also make different predictions about body shape. The Purcell inefficiency is relatively insensitive to differences in body shape, especially far from the wall. However, based on the small differences ($\approx 8\%$), the optimal body far from the boundary is short and thick, whereas the optimal body near the wall is short and thin. The Purcell inefficiency is the only quantity that makes different predictions about the optimal body near and far from the boundary. The Purcell inefficiency also predicts that bacterial motility systems become generally more efficient near the boundary, which would suggest a natural benefit for all bacteria to move near boundaries that are independent of any other biologically relevant activities. 

Unlike the Purcell inefficiency, the energy cost per distance traveled and the energy cost per body mass per distance traveled (metabolic energy cost) both predict larger energy costs for moving near a surface. However, they make opposite predictions about the optimal body size. The energy cost per distance suggests short and thin cell bodies are most efficient, and the metabolic energy cost suggests long and thick cell bodies are most efficient. Though increasing body size results in a greater energy cost for moving a given distance, the increase in body size results in a smaller relative energy expenditure. The energy per distance predicts the same optimal body as predicted by the fastest swimming speed, and metabolic energy cost predicts the same optimal body as predicted by motor rotation rate. Only the Purcell efficiency predicts a short, thick body is optimal, and this occurs only far from the boundary.

Although the Purcell efficiency has been a popular quantity of analysis, we believe it has several important shortcomings that warrant discussion, at least one of which was anticipated by Purcell. 
First, the Purcell efficiency is dependent only on the geometry of the body and flagellum and not on the motor's torque-speed response characteristics. 
From a physical standpoint, it is interesting to find such an invariant quantity, but from a biological standpoint,
it does not assess the bacterial motility system's thermodynamic efficiency because it ignores motor mechanics.

Second, the Purcell efficiency is defined to be the ratio between the minimum power required to translate the cell body and the power actually dissipated during the bacterial motion. In our simulations, we find the maximal efficiency is in the range of 1-2\%, similar to what others have found \cite{purcell1977life, chattopadhyay2006swimming, he2014propulsive}. These two quantities (the minimum power vs the actual power) are clearly of very different orders, which suggests that least power needed may not be an appropriate reference quantity.To give a biophysical interpretation to the least power needed to translate the cell body, some authors have suggested that it represents the ``useful'' portion of the power dissipated during motion,
\cite{chattopadhyay2006swimming,acemoglu2014effects} but we believe this is a misconception. The bacterium is non-inertial; therefore, the force acting on the cell body by the fluid is exactly balanced by the force acting on the flagellum by the fluid (assuming no net body forces). Both the bacterial body and the flagellum have the same axial velocity (in a rigid model); therefore the power dissipated due to the axial fluid drag on the body is exactly compensated by the power input by the axial fluid force exerted on the flagellum.

Finally, as Purcell noted in 1977, the efficiency of the bacterial motility system is probably best characterized by the energy consumption relative to the overall metabolic budget of the organism \cite{purcell1977life}. This suggestion led us to consider the metabolic energy cost introduced in this paper. The actual amount of that metabolic budget used for motility is a small fraction, which led Purcell \cite{purcell1977life} to suggest that bacterial motility is not really subject to strong selective forces toward optimal efficiency. Our data do not say whether evolutionary processes tend to minimize the energy cost of bacterial motility, but a plausible counterargument is that the bacterium needs to consume most of its energy for other biological functions and has only a small fraction available for motility.
Thus, small absolute changes in energy consumption correspond to large relative changes in the energy available for motility, resulting in a significant selective pressure to make the motility system as efficient as possible. Many research questions about how physical interactions between bacteria and their environment result in selective pressures in evolutionary processes remain open, despite significant progress in the field. Modern computational simulations and methods such as MRS and MIRS will remain important tools for quantifying microscopic bacterial motion with precision. In this work, we presented for the first time in the literature a procedure for calibrating MRS and MIRS using macroscopic dynamically similar experiments. Calibrating models in this way helps to ensure simulations give accurate quantitative results.
In future work, we will extend the macroscopic experimental system to consider a wider variety of possible geometries relevant to bacterial motility and make comparisons with biological measurements. 
\clearpage

\textbf{Funding:  }This research was partially funded by NSF DMS-1720323 to H.N. and N.C., and NSF MRI-1531594 to H.N. We thank Trinity University for the Summer Research Grant to O.S. and the provision of computational resources. We would also like to thank the Faculty Development Fund at Centre College for research support to B.R.

\acknowledgments{We wish to thank undergraduate students Asha Ari, Alexandra Boardman, Tanner May and Mackenzie Conkling and Prof. Philip Lockett for their assistance in collecting experimental data at Centre College. We also acknowledge the contributions of Mica Jarocki and David Clark at Trinity University to the initial implementation of the model bacterium. We would also like to thank Deon Lee for her support by editing the manuscript.}

\bibliography{mainNotes.bib}

\end{document}